\documentclass[pra,twocolumn,superscriptaddress,showpacs]{revtex4}

\usepackage{graphicx,color}
\usepackage{amsmath}
\usepackage{amssymb}

\renewcommand{\Re}{\mathop{\text{Re}}\nolimits}

\renewcommand{\Im}{\mathop{\text{Im}}\nolimits}

\newcommand{\Tr}{\mathop{\text{Tr}}\nolimits}
\newcommand{\ket}[1]{|{#1}\rangle}

\newcommand{\bracket}[2]{\langle#1|#2\rangle}

\newcommand{\beq}{\begin{equation}}
\newcommand{\eeq}{\end{equation}}

\newcommand\bwt         {\begin{widetext}}
\newcommand\ewt         {\end{widetext}}

\begin{document}

\title{Topological phase transitions driven by non-Abelian gauge potentials in optical square lattices}

\author{M. Burrello}
\affiliation{Instituut-Lorentz, Universiteit Leiden, P.O. Box 9506, 2300 RA Leiden, The Netherlands}

\author{I. C. Fulga}
\affiliation{Instituut-Lorentz, Universiteit Leiden, P.O. Box 9506, 2300 RA Leiden, The Netherlands}

\author{E. Alba}
\affiliation{Instituto de F\'{i}sica Fundamental, IFF-CSIC, Calle Serrano 113b, 28006 Madrid, Spain}

\author{L. Lepori}
\affiliation{IPCMS (UMR 7504) and ISIS (UMR 7006), Universit\'{e} de Strasbourg and CNRS, Strasbourg, France}
\affiliation{Departamento de F\'{i}sica, Universitat Aut\`{o}noma de Barcelona, E-08193 Bellaterra, Spain}

\author{A. Trombettoni}
\affiliation{CNR-IOM DEMOCRITOS Simulation Center, Via Bonomea 265, I-34136 Trieste, Italy}
\affiliation{SISSA and INFN, Sezione di Trieste,
via Bonomea 265, I-34136 Trieste, Italy}

\begin{abstract}
We analyze a tight-binding model of ultracold fermions loaded in an optical square lattice and 
subjected to a synthetic non-Abelian gauge potential featuring both a magnetic field 
and a translationally invariant SU(2) term. 
We consider in particular the effect of broken time-reversal symmetry and its role 
in driving non-trivial topological phase transitions.  
By varying the spin-orbit coupling parameters, 
we find both a semimetal/insulator phase transition and a topological phase transition between 
insulating phases with different numbers of edge states. The spin is not a conserved quantity of the system and the topological phase transitions can be detected by analyzing its polarization in time of flight images, 
providing a clear diagnostic for the characterization of the topological phases through the partial entanglement between spin and lattice degrees of freedom.
\end{abstract}



\maketitle

\section{Introduction}
The growing interest in topological phases of matter calls for an extension of experimental pursuits and realizations of topologically non-trivial systems to a broad variety of setups, in order to investigate the universal nature of these phases and exploit their remarkable properties. 
The recent major advances in identifying topological phenomena in solid state systems \cite{hasankane,zhang11} motivate the implementation and the study of new topological models in ultracold atomic systems. The rationale behind such efforts is that in these systems it is possible to simulate general classes of (Abelian and non-Abelian) synthetic gauge potentials \cite{dalibard11,Lewensteinbook}: this possibility 
is crucial in inducing topological phases, with a potentially high control of 
the physical parameters governing their phase diagrams \cite{goldman10}.

An important tool available in ultracold atomic setups is provided by the control of the dimensionality and the geometry of superimposed optical lattices. Such a control has been recently used in an experimental implementation of honeycomb lattices for ultracold fermions \cite{Tarr}. 
Several schemes have been discussed in the literature \cite{alba11,goldman13} with the aim of engineering topologically non-trivial two-dimensional systems 
on honeycomb lattices, which naturally reproduce well-known paradigmatic models such as the Haldane model \cite{haldane}. The discussed setups rely on the laser implementation of tight-binding models on honeycombs having both nearest-neighbor and next-nearest-neighbor hoppings: these hoppings simulate the presence of a staggered magnetic field which breaks time-reversal symmetry and opens band gaps. In this way one can obtain different topological phases, usually corresponding to topological insulators or semimetals, which mimic the quantum Hall physics and present different patterns of protected edge states. 

The key property of the honeycomb lattice, namely to have Dirac points, is also exhibited by the $\pi$-flux square lattice \cite{affleck,bernevig,manes}, a tight-binding model with only nearest-neighbor hoppings on a square lattice in a perpendicular magnetic field having flux $\pi$. The structure of Dirac points is unaltered when passing from a two-dimensional square to a three-dimensional cube with $\pi$-flux for each plaquette \cite{hasegawa,laughlin,LMT}. In two dimensions, one can explicitly see the 
similarity between the tight-binding models on the honeycomb and the $\pi$-flux square lattices by considering a continuous path smoothly interpolating 
between these two lattices and showing that the Dirac points are well defined across this path \cite{MLT}. 

In Ref.~\cite{lewenstein13} it has been shown that the anomalous quantum Hall regime exemplified by the Haldane model does not necessarily require the honeycomb geometry to be realized. It is possible to obtain equivalent Hamiltonians with the same topological properties by considering systems with only nearest-neighbors hoppings, designed on a square lattice with $0$ or $\pi$-flux. To reach non-trivial topological phases one has to introduce in the square lattice an inner degree of freedom and thus use a two-component Fermi mixture in a non-Abelian gauge potential \cite{osterloh05,goldman09a,goldman09b}. 
The tunneling terms along the two directions of the square lattice involve non-trivial rotations of the atomic pseudospin degree of freedom and in general do not commute with each other. Therefore, the non-Abelian gauge potential can be compared to spin-orbit coupling affecting the dynamics of the atoms. A spin-orbit term in the Hamiltonian not only allows us
to reproduce the anomalous quantum Hall regime, but also constitutes a key element to go beyond the quantum Hall symmetry class. This fact is clearly discussed in \cite{goldman12,morais,morais2}, where  a system obtained from the Kane-Mele model through the addition of an external magnetic field is analyzed. This system is characterized by a topological phase transition from a (weak) quantum spin Hall regime to a quantum Hall regime. Such a transition 
is related to the breaking of time-reversal symmetry and is signaled by the fact that, by varying the ratio between the spin-orbit coupling and the magnetic field terms, the edge structure undergoes a transformation: in the quantum spin Hall regime the system is characterized by pairs of counter-propagating edges, whereas in the quantum Hall phases only chiral edge modes appear.

A physical realization of these and related tight-binding models could be obtained by superimposing optical lattices to ultracold gases in synthetic gauge potentials implemented using spatially dependent optical couplings between internal states of the atoms as in Ref.~\cite{lin09}. However, since a $\pi$-flux per plaquette is needed, this implementation could be rather challenging. The same tight-binding models can be also obtained relying on the construction presented in Ref.~\cite{jaksch03}, which associates non-trivial phases to the tunneling amplitudes between nearest-neighbor and next-nearest neighbor-sites through the interaction of two atomic 
hyperfine species with counter-propagating Raman beams. These setups can efficiently reproduce spin-orbit terms to simulate topological insulators \cite{mazza12,lewenstein13}. Experimental results along this line have been recently obtained: a tunable effective magnetic field for ultracold atoms was implemented using Raman-assisted tunneling in an optical superlattice \cite{aidelsburger11}, while a one-dimensional chain with complex tunneling matrix elements was obtained from a combination of Raman coupling
and radio-frequency magnetic fields \cite{jimenez12}. Strong non-Abelian gauge potentials were created using shaken spin-dependent square lattices \cite{hauke12}.

In two-dimensional solid state systems the detection of non-trivial topological phases is usually based on transport measurements of charge and spin along the protected edge modes. Conductance measurements are however challenging in atomic gases \cite{brantut12}, and other tools can be adopted to detect the presence of edge modes \cite{goldman13b,goldman13d}. Apart from edge mode detection, bulk properties can also be exploited to investigate topological phases. The main examples are provided by the analysis of the Hall response of ultracold gases, which allows the detection of the Chern number of the filled bands through the observation of the center of mass motion under an external force \cite{leblanc12,barberan13,dauphin13}, and by the observation of the Bloch oscillations, which may permit us to detect 
the Berry connection of the spectral bands of lattice systems \cite{cooper12,lee13}. 

Several techniques have been proposed so far to relate the Chern number of topological insulators in ultracold gases to time of flight images \cite{alba11,goldman13,satija11,troyer13}. In particular, in Ref.~\cite{alba11} it has been shown that time of flight measurements, resolved with respect to the different spin species, allow us to reconstruct the spin polarization of the lowest band in the first Brillouin zone. From this polarization one can calculate a spin winding number which, for the Chern insulators related to the Haldane model, directly provides the Chern number of the occupied band, and thus the number of protected edge states \cite{goldman13}. Analyzing the polarization may permit us to test the topological characteristics of a system starting from bulk properties that can be easily detected in current experiments.

The Haldane model and its ultracold atom implementations or extensions are characterized by Hamiltonians which can be expressed in terms of a two-site basis cell. In the particular systems studied in \cite{alba11,goldman13,lewenstein13}, this allows an exact mapping between the Chern number of the spectral bands and the winding number of the spin polarization in the Brillouin zone. The goal of this paper is instead to explore a more general case where the direct connection between the Chern number of the spectral bands and the related spin winding number is no longer present.

To this purpose we consider a square lattice, with nearest-neighbor hoppings only, where atoms of two different species are subjected to both a magnetic field (equal for both species) and an SU(2) gauge potential which is translationally invariant \cite{goldman09b}. The general model studied here encompasses different interesting limiting cases and it is convenient for the study of topological phase transitions driven by non-Abelian gauge potentials. Denoting by $q$ the strength of the non-Abelian part of the gauge potential and by $B$ the magnetic field, for $q=0$ one can retrieve for a flux $\Phi_0/2$ the $\pi$-flux square lattice and its Dirac points \cite{affleck}, while for $B=0$ one simply has a spin-orbit coupled 
lattice model. In the continuous limit (i.e., without the lattice potential), for $q=0$ the structure of (doubly degenerate) Landau levels is in turn retrieved and the addition of the non-Abelian term in the gauge potential does not break the Landau levels \cite{burrello10,burrello11}, giving rise to non trivial fractional quantum Hall states extensively studied in bosonic gases \cite{palmer11,grass12,komineas12,grass13}. In the lattice, a fractional magnetic flux per plaquette $\Phi_0/n$ creates, in general, $n$ bands that are further split by the non-Abelian tunneling terms. 

We show in the following that, due to the effect of the non-Abelian potential, non-trivial topological phases arise. We classify them 
according to the standard classification of topological insulators and superconductors \cite{hasankane,ludwig08}. As a function of $q$ the system undergoes phase transitions between a topological semimetal regime and two 
different topological insulating phases. We consider in particular the effect of time-reversal symmetry breaking, comparing results with the ones obtained for $n=2$ (with time-reversal symmetry). Finally, we show that also in this more general framework, with a partial entanglement between spin and lattice degrees of freedom, the spin winding number still provides useful information about the phases of the system and may allow for their experimental detection. 

\section{Description of the model}
We consider a tight-binding model of two-component fermionic atoms loaded in a square lattice. We assume that the atoms are subjected to both a synthetic magnetic field, with a flux $\Phi$ per plaquette, and a translationally invariant non-Abelian SU(2) gauge field, which plays the role of spin-orbit coupling. These two elements influence the nearest neighbor tunneling amplitudes in different ways along the $x$ and $y$ directions, with their effect being described through the following U(2) gauge potential \cite{goldman09b,palmer11}:
\begin{equation} \label{potgen}
 \vec{A} = \frac {2 \pi \, \Phi}{a^2 \, \Phi_0} \left(0,x\right) 
\sigma_0 + \frac{q}{a}\left(\sigma_x,\sigma_y \right) \, ,    
\end{equation}
where $\Phi_0$ is the elementary flux of the synthetic magnetic field, 
corresponding to an Abelian phase $2\pi$ acquired by an atom surrounding a plaquette, $a$ is the lattice constant, $q\in\left[0,\pi \right]$ characterizes the intensity of the non-Abelian component, $\sigma_x,\sigma_y$ are the usual Pauli matrices and $\sigma_0$ is the $2 \times 2$ identity matrix. We consider only positive values of $q$, since the case with $q<0$ is equivalent up to a basis transformation, $\vec{A}(-q) = \sigma_z \, \vec{A}(q) \, \sigma_z$. The first term in Eq.~\eqref{potgen} defines the Abelian contribution of the gauge potential in the Landau gauge, where only the tunnelings along $y$ assume a position-dependent phase, not depending on the spin degree of freedom. The second term describes instead the spin-dependent non-Abelian term, which is translationally invariant and it is gauge equivalent to both a Rashba and a Dresselhaus spin-orbit term.

\begin{figure}[ht]
\centering
\includegraphics[width=7cm]{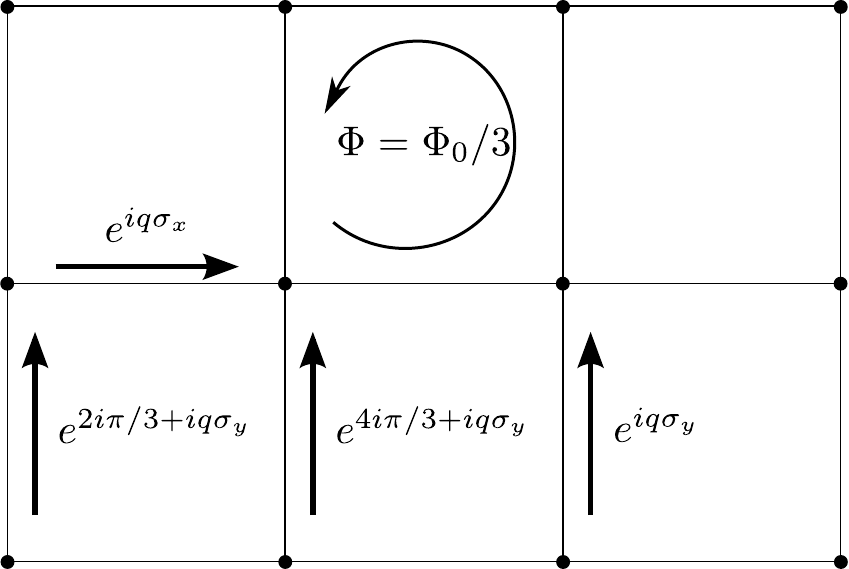}
\caption{Square lattice model with non-Abelian gauge potential. The tunneling operators $U_x$ and $U_y$, defined in Eqs. (\ref{ux},\ref{uy}), are represented for $\Phi=\Phi_0/3$. $U_x=e^{iq\sigma_x}$ is independent on the position, whereas $U_y$ is characterized by a phase dependent on $x$ which determines the Abelian flux $\Phi$ per plaquette.}
\label{lattice_fig}
\end{figure}

The tight-binding Hamiltonian of the system reads:
\begin{equation} \label{hamgen}
 H = t\sum_{\vec{r}, s, s'} \left[ U_{x,ss'} \, c^\dag_{\vec{r}+\hat{x},s'} \, c_{\vec{r},s} + U_{y,ss'}(x) \, 
c^\dag_{\vec{r} + \hat{y},s'}c_{\vec{r},s} \right] + \, \mathrm{h.c.} \, ,
\end{equation}
where $\vec{r}=(x,y)$ denote the lattice sites, $s$ labels the two components and 
the tunneling matrices $U_x$ and $U_y$ are defined as:
\begin{align}
 &U_x = \exp\left( {i\int_{x,y}^{x+a,y}A_x \, \mathrm{d}x}\right) = \exp\left( {iq \, \sigma_x}\right) \,, \label{ux} \\
 &U_y = \exp\left( {i\int_{x,y}^{x,y+a}A_y \, \mathrm{d}y}\right) =\exp\left( {i 2\pi \frac{x}{a} \frac{\Phi}{\Phi_0} + iq \, \sigma_y}\right) \, . \label{uy} 
\end{align}
Following our gauge choice \eqref{potgen}, only $U_y$ depends on position through the Abelian phase $2\pi \frac{x}{a} \frac{\Phi}{\Phi_0}$ (see Fig. \ref{lattice_fig}).

We observe that the Hamiltonian \eqref{hamgen} is invariant under the 
exchanges $t \to -t$, $q \to \pi-q$ through the basis transformation $\sigma_z$. This implies that the sign of $t$ can be chosen without loss of generality (of course, if $t \to -t$, then the following results are valid provided that $\nu \to 2-\nu$, where $\nu$ is the number of fermions per site). The validity of the Peierls substitution leading to the tight-binding 
Hamiltonian \eqref{hamgen} has been discussed for spin-orbit coupled atoms in optical lattices \cite{radic12}, where it was shown that for large enough values of the strength of the periodic potentials it 
agrees with results obtained from numerically computed Wannier functions.

To understand the nature of the gauge potential \eqref{potgen} it is useful to calculate the Wilson loop around a plaquette \cite{goldman09b}: the Abelian part of the potential gives rise to the phase $2\pi \, \Phi / \Phi_0$, whereas the non-Abelian term, due to the commutation relation between $U_x$ and $U_y$, generates a non-trivial unitary operator for each $q \ne 0 , \pi$. However, such a plaquette operator does not depend on its position, emphasizing the translational symmetry of the system along $\hat{y}$. In particular, if the magnetic flux is commensurate with the elementary flux, $\Phi/\Phi_0 = m/n$, then the wave function can be written as 
\beq
 \psi(x,y)=e^{ik_x x} \, e^{ik_y y} \, u (x) \, ,
\eeq
where $u (x)$ is periodic with period $n$ (hereafter we set $a=1$). 
The resulting first Brillouin zone is given by $k_x \in \left[ 0,2\pi/n\right) $ and $k_y \in  \left[ 0,2\pi\right)$, as in the case when only the magnetic field is present \cite{LandauStat2}. The parameter $q$ does not influence the definition of the Brillouin zone and of the allowed quasi-momenta since the non-Abelian part of the gauge potential is translationally invariant.

We conclude this Section with a brief analysis of possible implementations of the potential \eqref{potgen}. There are two different ingredients to be combined: an Abelian net flux per plaquette (parametrized by $\Phi$) and a spin-dependent hopping element (parametrized by $q\vec{\sigma}$). A detailed discussion of the simulation of non-Abelian SU(2) potentials with alkaline atoms can be found in Ref.~\cite{mazza12}. There, the spin degree of freedom is coded in two hyperfine levels of $^{40}{\rm K}$. Zeeman splitting allows for independent addressing of the transitions between different spin states, thus creating the desired spin-dependent hopping amplitude. An additional superlattice could then be used to engineer a rectified magnetic flux, independent of the spin state of the system \cite{gerbier10}. 

A comprehensive implementation proposal for both the Abelian and non-Abelian potential can be found in \cite{lewenstein13}. This setup features a state-dependent lattice loaded with the ground state and a metastable excited state of $^{171}{\rm Yb}$. The narrow transition between the two trapped states guarantees a negligible heating rate, and naturally paves the way for implementing a rectified flux scheme. Polarization of the Raman lasers is used to engineer tunneling elements between different spin states, whereas a superlattice is again imposed to distinguish resonances in the two spatial directions.

As a final note, we would like to mention that the measurement protocol outlined in Section~\ref{Detection} suits the presented implementation proposals; in particular, the time of-flight measurement of the pseudospin degree of freedom is simplified by the presence of the magnetic field, which discriminates between different spin states.

\section{Topological phases and topological phase transitions}

\subsection{Conserving time-reversal symmetry: magnetic flux $\Phi_0/2$} 
The first case we analyze is the one with $n=2$, where an atom acquires an 
Abelian phase $\pi$ encircling a plaquette. This value of the flux is the only one maintaining time-reversal symmetry. The potential \eqref{potgen} reads:
\begin{equation}
 \vec{A}={\pi}(0,x) \, \sigma_0 +q\left(\sigma_x,\sigma_y \right) \, . 
\label{potpi}
\end{equation}
In this case the unit cell of the system is composed of two subsets of sites 
corresponding to even and odd $x$ coordinates. Therefore we can define an effective pseudospin $1/2$ degree of freedom and a new set of Pauli matrices $\tau_i$ referring to it, with  $\tau_z=\pm1$ indicating even and odd $x$-coordinates respectively. 
The Hamiltonian \eqref{hamgen} with potential \eqref{potpi} reads in quasi-momentum space: 
\begin{multline}
 \frac{H(\vec{k})}{2t} = \cos q \cos k_y \, \tau_z \sigma_0 - \sin q \sin k_y \,  \tau_z \sigma_y + \\ + \cos q \cos k_x \, \tau_x \sigma_0 - \sin q \sin k_x \, \tau_x \sigma_x \, .
 \label{2Dpi}
\end{multline}
It is a $4 \times 4$ matrix involving direct products of Pauli matrices (i.e., $\tau_z \sigma_y$ stands for $\tau_z \otimes \sigma_y$ and so on). 
The Hamiltonian \eqref{hamgen} is expressed in terms of \eqref{2Dpi} as
\begin{equation}
 H=\sum_{\vec{k}, s, s', \tau, \tau'} c_{s',\tau'}^\dag(\vec{k}) \, H_{s'\tau',s\tau}(\vec{k}) 
\, c_{s,\tau}(\vec{k}) \, , 
\label{sum_H}
\end{equation}
where the sum is on the first Brillouin zone, $c_{s\tau}(\vec{k})$ is the Fourier transform of 
$c_{s\tau}(\vec{r})$ and $\tau=\pm1$ is the pseudospin index.

In order to understand the topological characteristics of the model, it is useful to examine the topological symmetry class in the Altland and Zirnbauer classification \cite{altlandzirnbauer,kitaev09,ludwig10}. We consider the set of discrete anti-unitary symmetries and we identify at half-filling ($\mu=0$) the following time-reversal symmetry $T$ and particle/hole-like symmetry $C$ \cite{ludwig08}:
\begin{eqnarray} \label{T_sym}
 \sigma_y \, H^T(\vec{k}) \, \sigma_y &= H(-\vec{k}) \quad {\rm T-symmetry}\\
 \tau_y  \sigma_y \, H^T(\vec{k}) \, \tau_y   \sigma_y &= -H(-\vec{k}) \quad {\rm C-symmetry};
\label{C_sym}
\end{eqnarray}
the corresponding chiral unitary symmetry is generated by generated by $\tau_y$.
Given these discrete symmetries, the topological symmetry class in the Altland and Zirnbauer classification \cite{altlandzirnbauer} is DIII for generic values of $q \ne p \, \pi/2$, with integer $p$. DIII is a topologically non-trivial class in two dimensions which can support helical Majorana edge modes \cite{helical}. For the particular values $q=0,\pi$ the gauge potential corresponds to the presence of the magnetic field only and the Hamiltonian 
acquires the additional SU(2) spin symmetry; for $q=\pi/2,3\pi/2$, instead, the two non-zero terms of the Hamiltonian, proportional to $\tau_z \sigma_y$ and $\tau_x \sigma_x$, commute and generate a U(1)$\times$U(1) symmetry, while $\sigma_z$ anti-commutes with the Hamiltonian. Finally, when the chemical potential $\mu \neq 0$ the T-symmetry \eqref{T_sym} is conserved, 
but the C-symmetry \eqref{C_sym} is broken, leading to a class AII system.

\begin{figure}[ht]
\centering
\includegraphics[width=7cm]{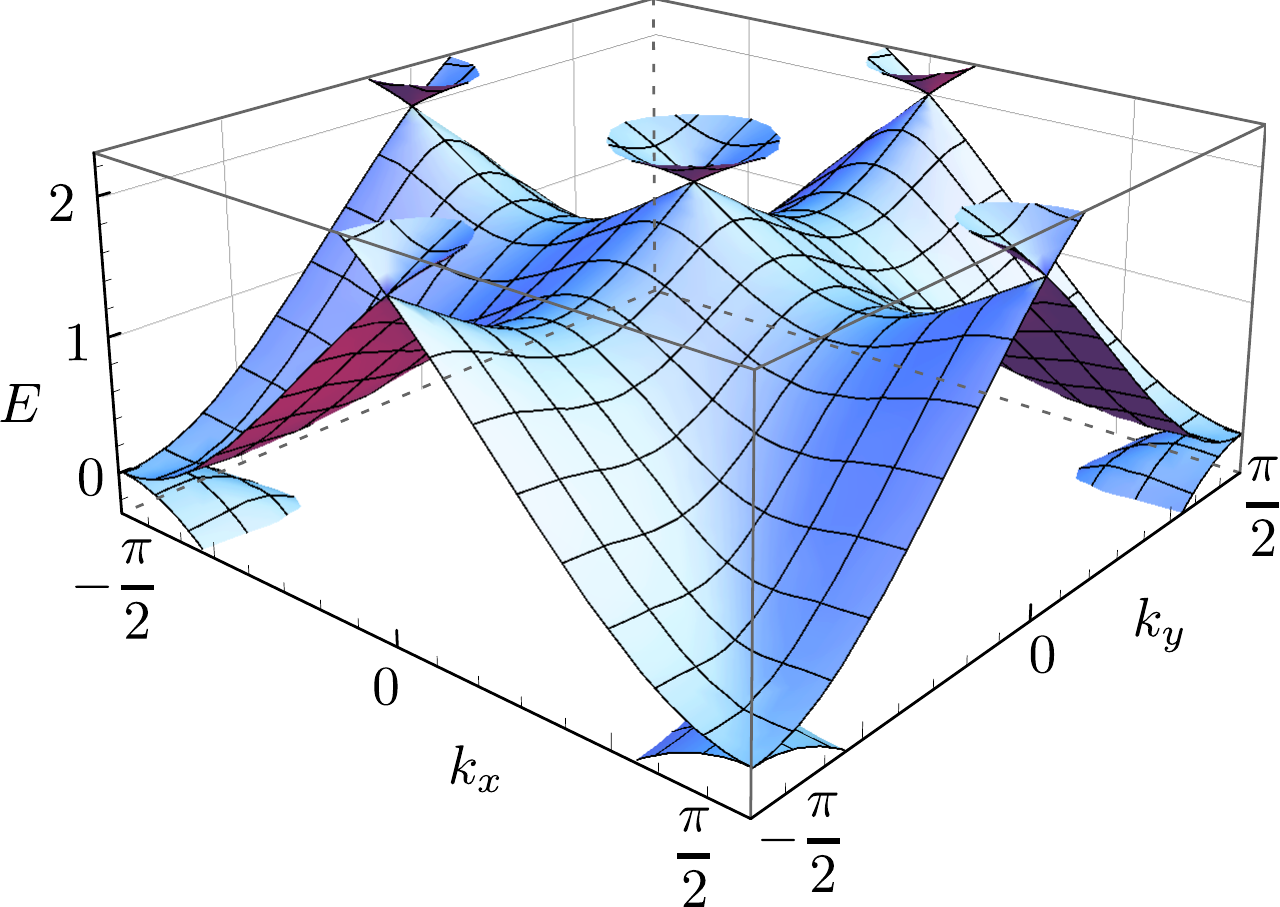}
\caption{(Color online). The spectrum of the Hamiltonian \eqref{2Dpi} is shown for $q=\pi/4$. The Dirac cones connecting the energy bands $3$ and $4$ are clearly visible in the upper part of the plot at $E=2$, whereas the gap between the bands $2$ and $3$ closes at $(k_x, k_y)=(\pm \pi/2,\pm \pi/2)$ and $E=0$. The spectrum is symmetric about $E=0$ and no energy gap appears for any value of $q$. The energy is expressed in units of $t$.}
\label{pibands}
\end{figure}

The effective realization of the topological insulating phase 
with Majorana edge modes depends critically on the band structure and 
on the appearance of energy gaps, as a function of the Hamiltonian parameter $q$. For generic values of $q$ the system has four energy eigenstates: 
we label them in increasing order ($\{1,2,3,4\}$). The corresponding bands in the Brillouin zone intersect for some particular quasi-momenta. Bands $1$ and $2$ coincide for every $q$ at $(k_x, k_y)=(0,0)$, with energy $E_-(0,0) = -2\sqrt{2} \, t\cos{q}$. The same happens for bands $3$ and $4$, with energy $E_+(0,0) =  2\sqrt{2} \, t\cos{q}$. Bands $2$ and $3$ instead have a common point, still for every $q$, at $(k_x, k_y)=(\pi/2, \pi/2)$, with vanishing energy (see Fig. \ref{pibands}). We conclude that the Hamiltonian does not have an energy gap for any value of $q$, and the system is always in a metallic or semimetallic phase, depending on whether the chemical potential falls in a bulk band or is set exactly at the intersections between them.

The same result is obtained by introducing an anisotropy in the non-Abelian potential 
by considering 
\begin{equation} \label{pot_gen_gen}
 \vec{A}={\pi}(0,x) \, \sigma_0 + \left(q_x\sigma_x,q_y\sigma_y \right) \, .
\end{equation}
This  generalization does not affect the symmetry class; the system remains in class DIII (or AII). Again the bands always touch at some points of the Brillouin zone; therefore the system is always in a metal or semimetal phase and there is no phase transition driven by the variation of the non-Abelian gauge parameters $q_x$ and $q_y$.

In order to obtain nontrivial topological phases, characterized by symmetry-protected edge modes, and 
topological phase transitions between them, it is necessary to open a gap between some bands. One possibility  is to break time-reversal symmetry. In the following we will achieve this by considering a system with magnetic flux $\Phi_0/3$ per plaquette.

\subsection{Breaking of time-reversal symmetry: magnetic flux $\Phi_0/3$}
By introducing in the system a fractional magnetic flux per plaquette $\Phi=m \, \Phi_0/n$ with $n\ne 2$, one obtains a series of bands which are separated by energy gaps, corresponding to the deformation of the Hofstadter butterfly spectrum due to the presence of the non-Abelian potential. Such deformations have already been analyzed for several potential configurations \cite{osterloh05,goldman07,goldman09b}. We exploit the presence of these gaps to generate and identify the different topological phases that appear with the potential \eqref{potgen}, whose transitions can be driven by the non-Abelian parameter $q$.

We focus in particular on the case with Abelian flux ${\Phi_0}/{3}$ ($n=3$), 
which is the simplest one breaking time-reversal symmetry and presenting topological phase transitions. The corresponding gauge potential reads:
\begin{equation} \label{pot3}
 \vec{A}=\frac{2\pi}{3}(0,x) \, \sigma_0 + {q}\left(\sigma_x,\sigma_y \right) \, .
\end{equation}
The unit cell is composed of three sites and it is useful to introduce a new 
pseudospin-1 degree of freedom, the related operator being labeled by $T_z$, 
to identify the $x$ coordinate modulo 3. The hopping operator in the $\hat{y}$ direction is diagonal in $T_z$, whereas the hopping along $\hat{x}$ is not. By introducing the two hopping matrices
\begin{equation}
T_x = \begin{pmatrix}0& 0& 1 \\ 1& 0 & 0 \\ 0& 1 & 0 \end{pmatrix}\, , \quad    T_z=\begin{pmatrix} e^{i\frac{2\pi}{3}} & 0 & 0 \\ 0 & e^{i\frac{4\pi}{3}} & 0 \\ 0 & 0 & 1 \end{pmatrix} \, ,
\end{equation}
the tunneling operators in \eqref{ux} and \eqref{uy} read:
\begin{equation}
U_x = T_x \, e^{iq\sigma_x}\,,\qquad U_y=T_z \, e^{iq\sigma_y} \, 
\end{equation}
(direct product of matrices is always intended). 
$U_x$ and $U_y$ are $6\times 6$ matrices acting on the space defined 
by the tensor product of the lattice pseudospin-1 and the inner spin-1/2 degrees of freedom. 

Since $n=3$, the first Brillouin zone is defined by the quasi-momenta 
$k_x \in \left[0,2\pi/3 \right)$ and $k_y \in\left[0,2\pi\right) $. In the 
quasi-momentum space the Hamiltonian reads:
\begin{multline}
H = -\mu \, \mathbb{I}_6 + T_x \, e^{iq\sigma_x} e^{ik_x} + T_x^\dag \,  e^{-iq\sigma_x}e^{-ik_x} + \\ + T_z \, e^{iq\sigma_y}e^{ik_y} + T_z^\dag \, e^{-iq\sigma_y}e^{-ik_y} \, ,
\end{multline}
where $t$ is set to 1, $\mu$ denotes the chemical potential which changes the filling, and $\mathbb{I}_6$ is a $6\times 6$ unit matrix. The Hamiltonian can be easily rewritten as:
\begin{multline} \label{ham3}
H = -\mu \, \mathbb{I}_6 + \left[ \cos q \left( T_x \, e^{i k_x} + T_z \, e^{ik_y}\right) + \right. \\ \left. +i\sin q \left( T_x \, \sigma_x e^{ik_x} + T_z \, \sigma_y e^{ik_y} \right)\right]  + {\rm h. c.} \, .
\end{multline}
For generic values of $q$ the six eigenstates of the Hamiltonian \eqref{ham3}, 
with energies $E_\lambda(\vec{k})$, describe five different spectral bands separated by energy gaps. There are five because the two central ones are always connected through Dirac cones. 

Even though the bands are separated by gaps for every value of the quasi-momenta, due to their strongly bent shapes, indirect overlaps (the minimum of the higher band is lower than the maximum of the lower band) appear between the two lowest-energy bands and between the two highest-energy bands for $q$ in a neighborhood of $0$ or $\pi$. In these cases the non-Abelian potential is not strong enough to open an insulating phase separating these pairs of bands and when the chemical potential is such that two different bands are partially filled, the system is in a semimetal phase \cite{goldman13}, despite the absence of Dirac cones.

For $q$ far enough from $0$ and $\pi$ instead, the semimetal phases disappear and are substituted by insulating phases. Therefore the chemical potential 
$\mu$ determines the usual alternation of metallic 
and insulating phases appearing in correspondence to all the gaps of the system. 

Other phase transitions, of a topological nature, appear for $q=\pi/3$ and $2\pi/3$ where the gaps between the two top bands and the two bottom bands directly close in Dirac cones. To understand the signature of this phase transition we analyze the topological features of the system. Since $T_x^\dag \ne T_x^*$, one cannot find anti-unitary operators corresponding to discrete T- and C- symmetries. The Hamiltonian \eqref{ham3} is then in the same topological class of the quantum Hall effect, class A \cite{note1} in the Altland and Zirnbauer classification.

The Hamiltonian $H$ describes a system which is topologically non-trivial and 
characterized by the presence of edge modes. They are localized on the 
boundary of the two-dimensional system and therefore can be observed only in bounded geometries, like the cylinder. Their energies interpolate between subsequent bands and their number is related to the topological invariants of the spectral bands. Moreover, due to the non-Abelian term, they do not present a fixed spin orientation; the non-Abelian potential in \eqref{potgen} is indeed equivalent to a Rashba spin-orbit coupling, which yields to a dependence of the spin polarization on the quasi-momentum.

Let us consider now values for $q$ and the chemical potential such 
that the system is in a bulk insulating phase with an integer number of atoms per unit cell. Then the number of observed edge states per boundary is given by the sum of the Chern numbers of the bands with an energy smaller than the Fermi energy. Given the $6$-component eigenfunctions $\psi_\lambda(\vec{k})$ of H (with $\lambda$ labeling the bands), their Chern number $C_{\lambda}$ is defined by \cite{hasankane,tknn}:
\begin{equation} \label{chern}
 \mathcal{C}_\lambda = \frac{i}{2\pi} \int_{\rm BZ} d^2 k \, \Big( \bracket{\partial_{k_{x}} \psi_\lambda}{\partial_{k_{y}} \psi_\lambda} - \bracket{\partial_{k_{y}} \psi_\lambda}{\partial_{k_{x}} \psi_\lambda} \Big) \, .
\end{equation}
The sign of $\mathcal{C}_\lambda$ refers to the direction of the propagation 
of the edge modes associated with $\psi_\lambda(\vec{k})$.

In order to examine how the numbers and the features of the edge modes change by varying $q$ (which is supposed to be in a range out of the semimetallic phase), let us label the Hamiltonian eigenfunctions from the lowest to the highest energy by an index $\lambda = -2,-1,0^+,0^-,1,2$. For generic values of $q$ only bands $0^+$ and $0^-$ are directly connected through Dirac cones whereas the other bands are separated by energy gaps, although often with the indirect energy overlaps discussed above that determine the presence of topological semimetal phases \cite{goldman13}.

\begin{figure}[ht]
\centering
\includegraphics[width=7cm]{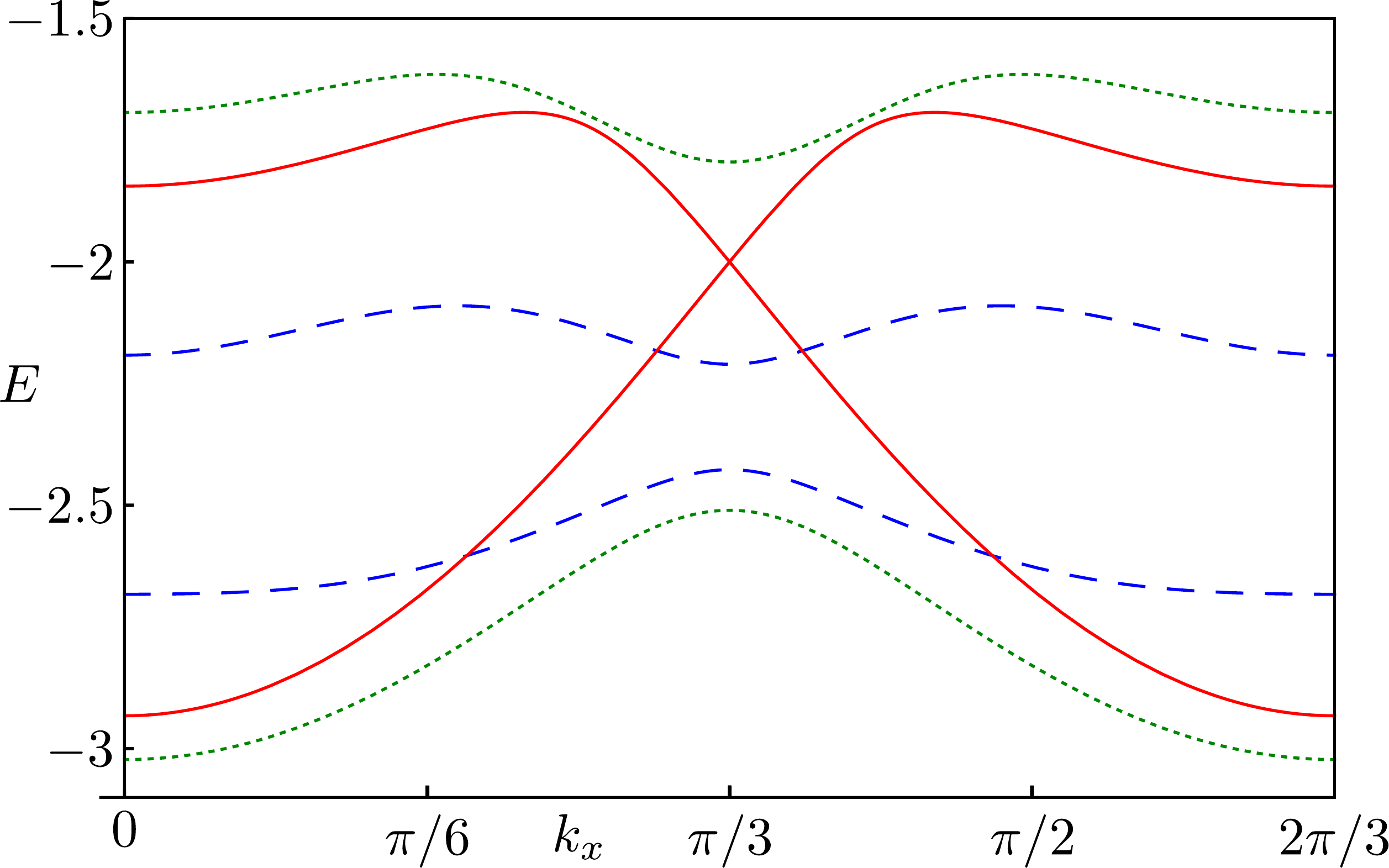}
\caption{(Color online). The energies of the two lowest states of the Hamiltonian \eqref{ham3} are plotted for $q=\pi/5$ (blue dashed line), $q=\pi/3$ (red line) and $q=2\pi/5$ (green dotted line), for $k_y=\pi/3$ and as a function of $k_x$. For $q=\pi/5$ and $q=2\pi/5$ the lowest bands are separeted by a gap and, for filling $\nu=1/3$, the system is in two different topological insulating phases. At $q=\pi/3$ the gap closes causing a topological phase transition. The energy is in units of the tunneling amplitude $t$ and the chemical potential is fixed at $\mu=0$.}
\label{transition1}
\end{figure}

The energy gap between the bands $-2$ and $-1$ closes at $q=\pi/3$ for $k_x=\pi/3$ and $k_y = \pi/3+ 2 \pi p/3$, with integer $p$ (see Fig.~\ref{transition1}). Analogously the gap between the two highest bands 
closes at the same value of $q$ for $k_x=0$ and $k_y= 2 \pi p/3$. 
The same gaps close also at $q=2\pi/3$, at interchanged quasi-momenta $\vec{k}$.

From the analysis of the spectrum it is clear that this closing of the gap 
corresponds to a further Dirac cone appearing in the bulk of the system. 
However this does not correspond simply to a crossing of Landau levels, 
as we would expect in a similar system in the continuous limit \cite{burrello10,burrello11}; 
instead, it is a topological phase transition which changes the Chern numbers of the involved bands, thus affecting the number of edge states in a geometry with boundaries. No change in the Chern number arises instead along the phase transition between a semimetal and an insulating phase.

To calculate the number of edge states, let us consider the spectrum of an infinite stripe with two edges. We discuss first the case in which only the lowest energy band is fully occupied, whereas all the others are empty (filling $1/3$, corresponding to one atom per unit cell). The number 
of propagating modes at each edge of the system coincides with the Chern number $\mathcal{C}_{-2}$, which is $+1$ for $0<q<\pi/3$ (or $2\pi/3<q<\pi$) and $-2$ for $\pi/3<q<2\pi/3$. Increasing the chemical potential and 
considering also the contribution of the second band $\mathcal{C}_{-1}$, 
we see from numerical calculations that the number of edge modes interpolating 
between the second and the lowest central band is $2$ for $0<q<\pi/3$ (or $2\pi/3<q<\pi$) and $-4$ for $\pi/3<q<2\pi/3$; this implies that $\mathcal{C}_{-1}$ also changes from $1$ to $-2$ and vice versa across the two phase transitions. Thus, going from $q<\pi/3$ to $q>\pi/3$, the edge states double and change their direction, as an effect of the non-Abelian potential. At $q=2\pi/3$ the transition has the opposite effect and brings the system back into the first insulating phase. An analogous behavior appears in the edge modes interpolating between the two highest energy bands, in the case of filling $5/3$.

These features have been numerically studied for the tight-binding Hamiltonian of the model using the Kwant software package \cite{kwant,kwant2} to evaluate the number of edge modes for an infinite stripe geometry of width 40 whose spectrum has been calculated as a function of $k_x$. The Chern numbers were instead independently calculated using a discretized version of Eq.~\eqref{chern} and dividing the Brillouin zone into $120 \times 120$ plaquettes.

The variation of the Chern numbers at the topological phase transition 
is similar to the one driven by the spin-orbit coupling generated by $\sigma_z$ in the honeycomb lattice model, as described in \cite{morais}. 
However, in the case analyzed here spin is not a conserved quantity, meaning that different spin species do not have separate Chern numbers. The phase transitions are driven instead by an off-diagonal Rashba coupling. 

\begin{figure}[htb]
 \centering
 \includegraphics[width=0.3\textwidth]{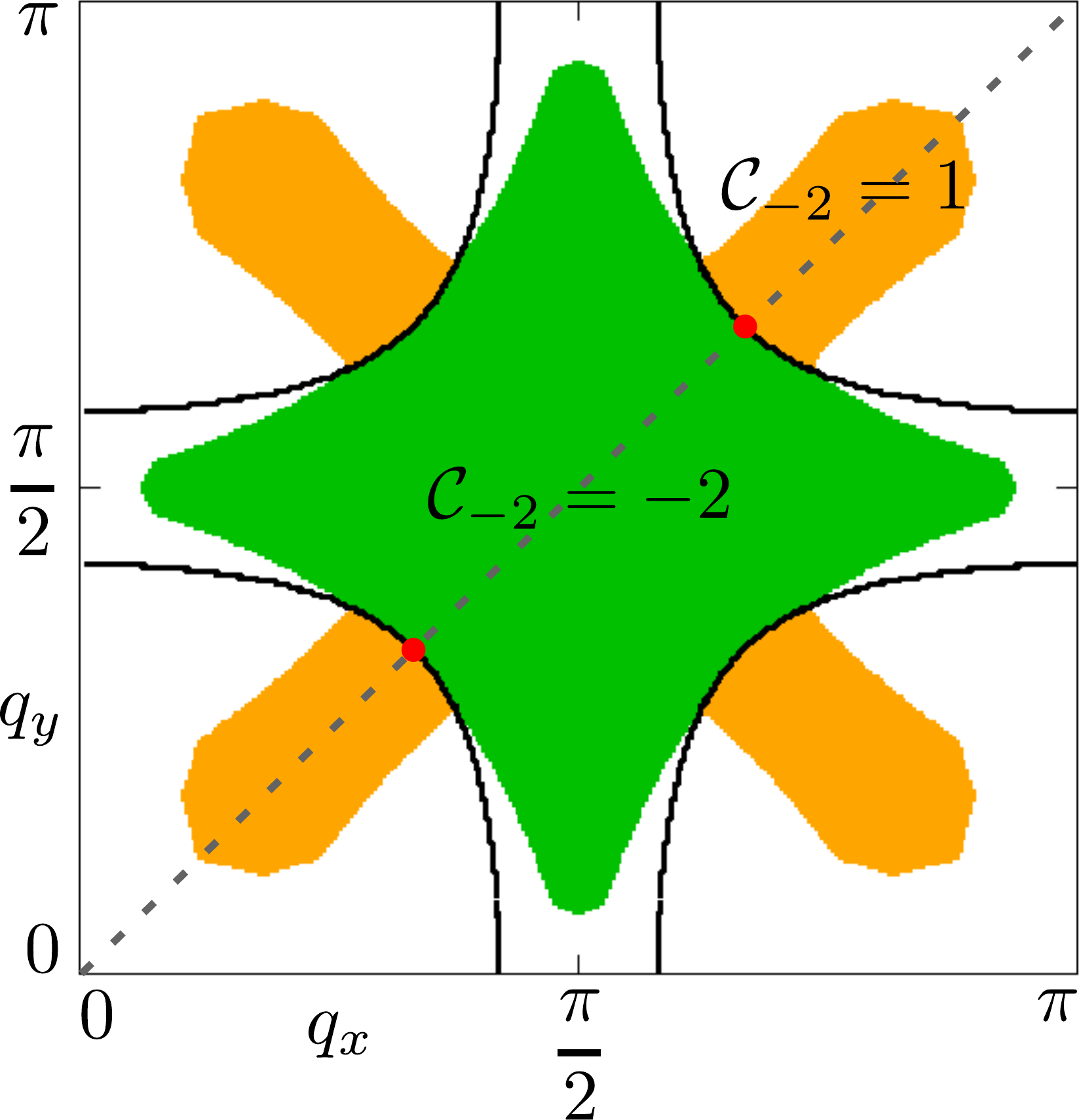}
 \caption{(Color online). Phase diagram at filling $1/3$ as a function of $q_x$ and $q_y$ for the gauge potential \eqref{pot_gen_gen_2}. The insulating regions characterized by Chern numbers ${\cal C}_{-2}=-2, 1$ are shown in green and orange respectively. Uncolored regions correspond to indirect overlap between bands (topological semimetal phase), whereas the black lines show the position of the Dirac cones corresponding to a discontinuity in the Chern number $\mathcal{C}_{-2}$. The red dots represent topological phase transitions at $q=\pi/3,2\pi/3$ along the diagonal $q_x=q_y=q$.} \label{fig:pd}
\end{figure}

Finally, the potential \eqref{pot3} can be generalized to the anisotropic case:
\begin{equation} \label{pot_gen_gen_2}
 \vec{A}=\frac{2\pi}{3}(0,x) \, \sigma_0 + \left(q_x \sigma_x,q_y \sigma_y \right) \, .
\end{equation}
The phase diagram of the anisotropic model at filling $1/3$ is shown in Fig.~\ref{fig:pd}. The $q_x = q_y$ pattern described before is shown by a gray dotted line, and the corresponding transition points between the two topological insulating phases are shown by two red dots.

For $q_x \neq q_y$ we still observe the presence of three different phases: 
a topological semimetal phase, characterized by a partial filling of the two lowest bands, and two distinct topological insulating phases, where the first band is completely filled and the second is empty. The insulating phases are characterized by a nonzero Chern number $\mathcal{C}_{-2}=1,-2$ respectively. 
Again the topological semimetal phase is distinguished by the topological 
insulating phases due to the presence of an indirect overlap of the bands; 
once this overlap disappears, we have a transition between these two regimes and the bulk conductance vanishes. The topological phase transition between the two insulating phases is instead related to a change in the Chern numbers, with parameters $q_x$ and $q_y$ determining the position of Dirac cones appearing in the spectrum at the transition.

\section{Detection of the topological phase transition}
\label{Detection}
Several techniques have been proposed to experimentally detect topological phase transitions like the one discussed in the previous Section. 
Some of them are based on the observation of the edge modes \cite{goldman13b,troyer13,goldman13d}, which in our model change both in number and in direction of propagation at the transition points, providing direct evidence for the topological transition. Other viable techniques rely on the possibility of detecting the dynamic response of the bulk under an external force or on other bulk properties \cite{cooper12,leblanc12,barberan13,dauphin13}. Finally, it has been pointed out in the literature that, at least in certain fermionic setups based on the Haldane honeycomb model \cite{haldane}, the time of flight images discriminating the inner spin-degree of freedom of the atoms allow a direct measurement of the Chern number of the lowest energy band, once the Fermi level is placed in the first energy gap \cite{alba11,goldman13}.

In the models analyzed in Refs.~\cite{alba11,goldman13} two triangular optical lattices, each trapping a different hyperfine state of one atomic species, 
are superimposed in order to obtain an overall honeycomb lattice model. 
The two spin species lay in the two sublattices and the models are characterized by a nearest-neighbor hopping which changes the spin species 
and a next nearest neighbor tunneling which preserves the spin. 
In such models the Hamiltonian can be written as a $2 \times 2$ matrix where the spin and sublattice degrees of freedom completely coincide: it can be shown \cite{tknn} that in the insulating phases the spin (Pontryagin) winding number in the first Brillouin zone coincides with the Chern number of the lowest band and thus with the number of edge states.

In the model considered here instead, $\Phi=\Phi_0/3$ and the Hamiltonian \eqref{ham3} is a $6 \times 6$ matrix with eigenfunctions that can be generically expressed as:
\begin{equation}
 \ket{\psi_\lambda(\vec{k})}=\sum_{\substack{s=\, \uparrow,\downarrow \\  x=0,1,2 }} 
c_{\lambda,sx}(\vec{k}) \, \ket{s}_{\rm spin} \, \ket{x}_{\rm lat} \, ,
\end{equation}
where the spin and lattice pseudospin (labeled above as lat) degrees of freedom do not coincide. The wave function $\psi_\lambda$ is described in the basis obtained by the tensor product of the spin and lattice pseudospin spaces and, due to the effect of the non-Abelian potential, $\psi_\lambda$ is not simply a direct product state of the form $\ket{\varphi_\lambda(\vec{k})}_{\rm spin} \otimes \ket{\chi_\lambda(\vec{k})}_{\rm lat}$, but it encodes some entanglement between the two degrees of freedom.

Despite the fact that the Hamiltonian does not define in a direct way 
a mapping from the Brillouin zone to the Bloch sphere, 
it is still possible to follow an approach similar to \cite{pachos13}
and define a spin winding number over the first Brillouin zone. 
This is because the spin is a periodic observable in the first Brillouin zone, and therefore a spin winding number can be properly defined and  experimentally measured through time of flight imaging \cite{alba11}. This spin winding number does not coincide with the Chern number defined in \eqref{chern}, because the latter requires in its definition a knowledge of the full wave function, not only of the spin part. Nevertheless we will show that the spin pattern of the wavefunctions in the different bands provides, in our model, enough information to detect the topological phase transition.

\begin{figure*}
 \includegraphics[width=0.3\textwidth]{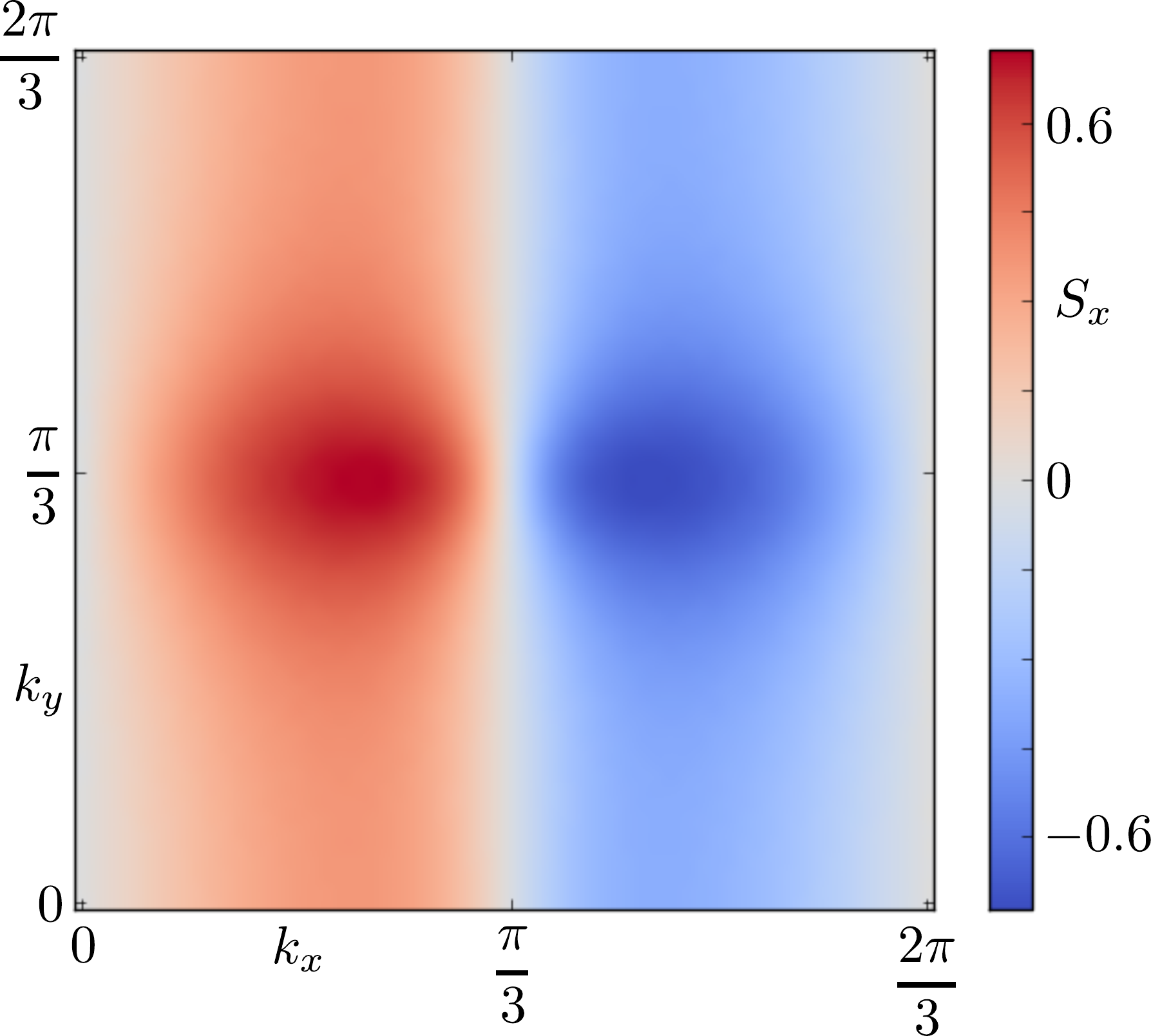}
 \includegraphics[width=0.3\textwidth]{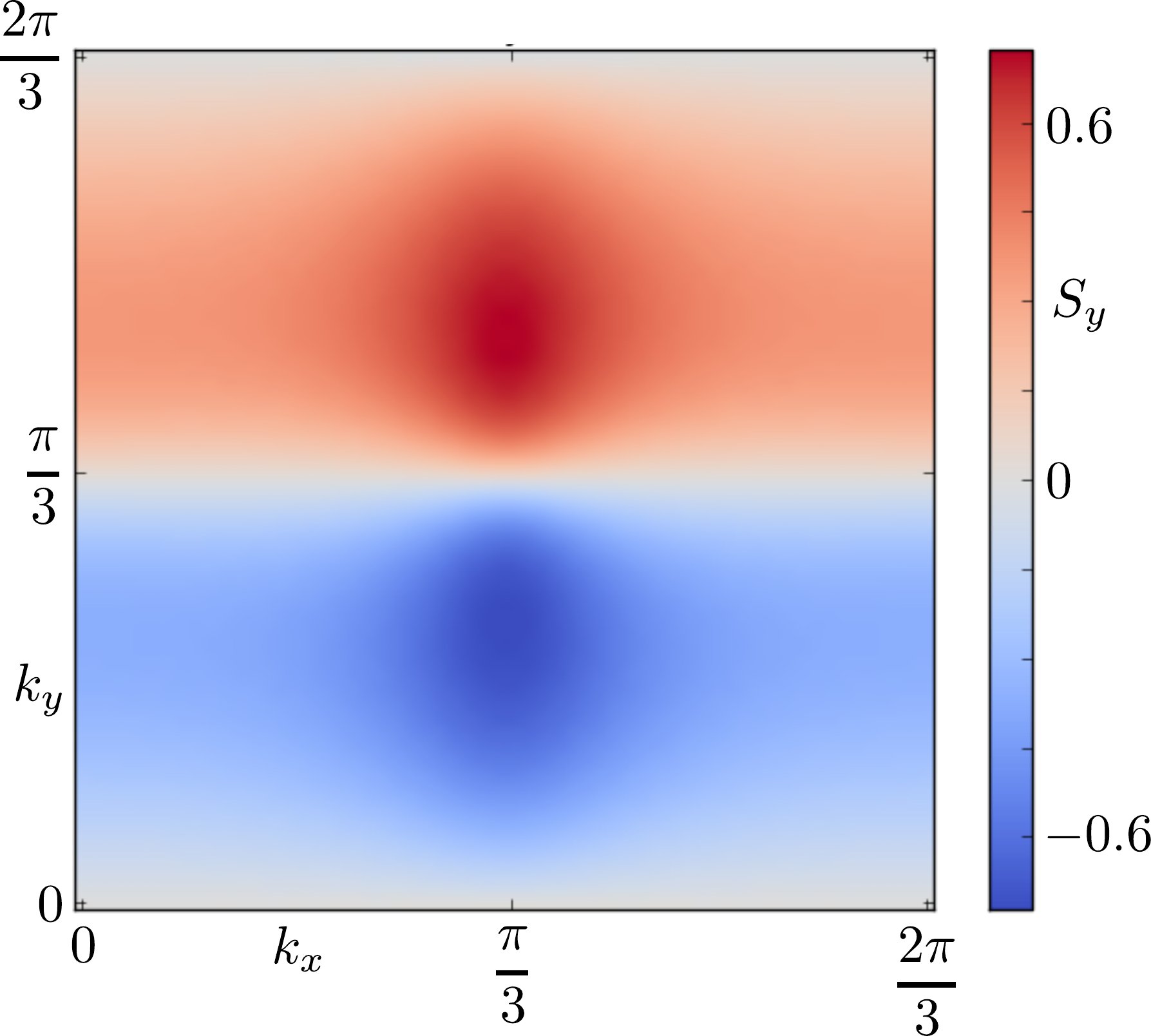}
 \includegraphics[width=0.3\textwidth]{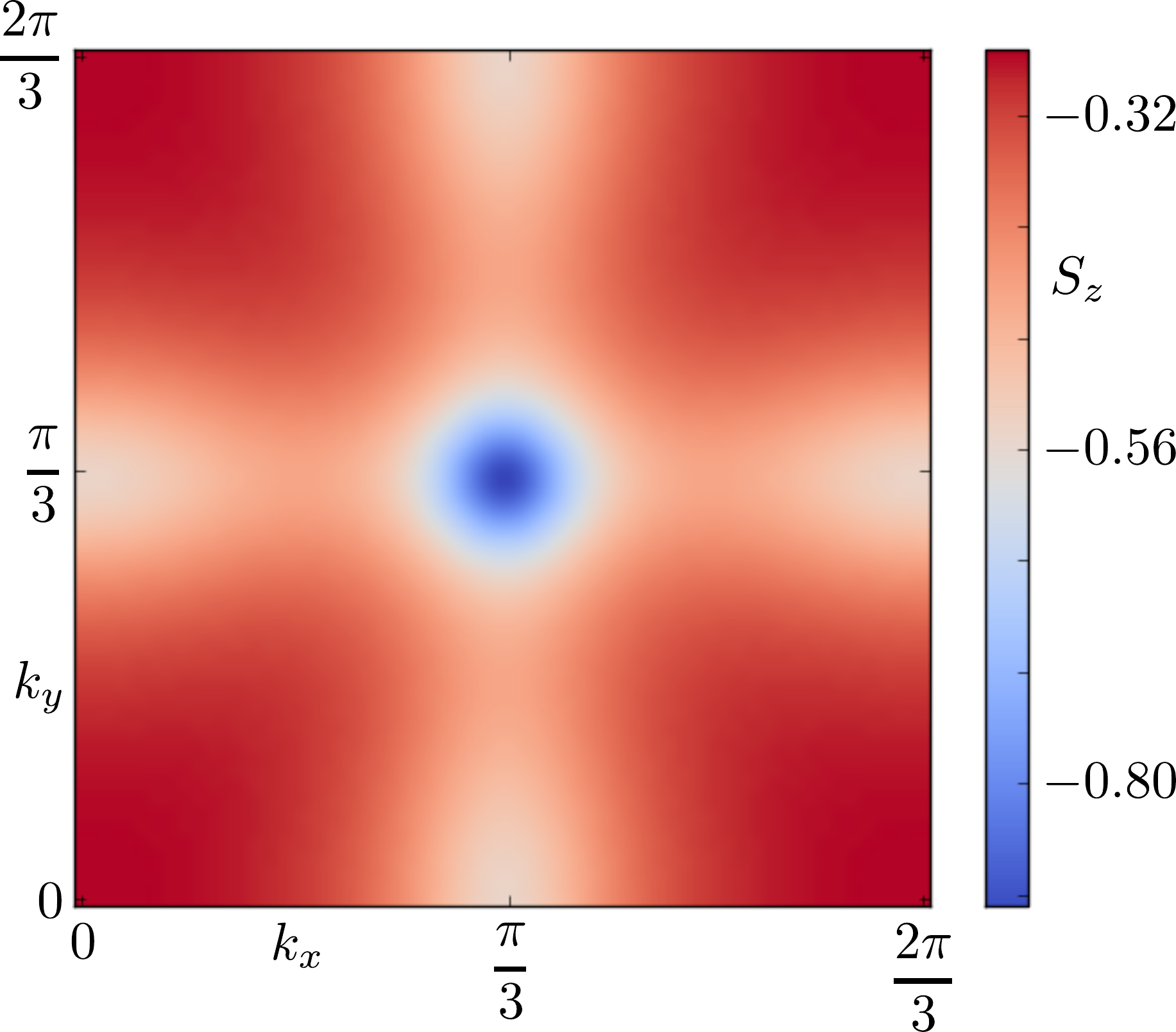}

 \includegraphics[width=0.3\textwidth]{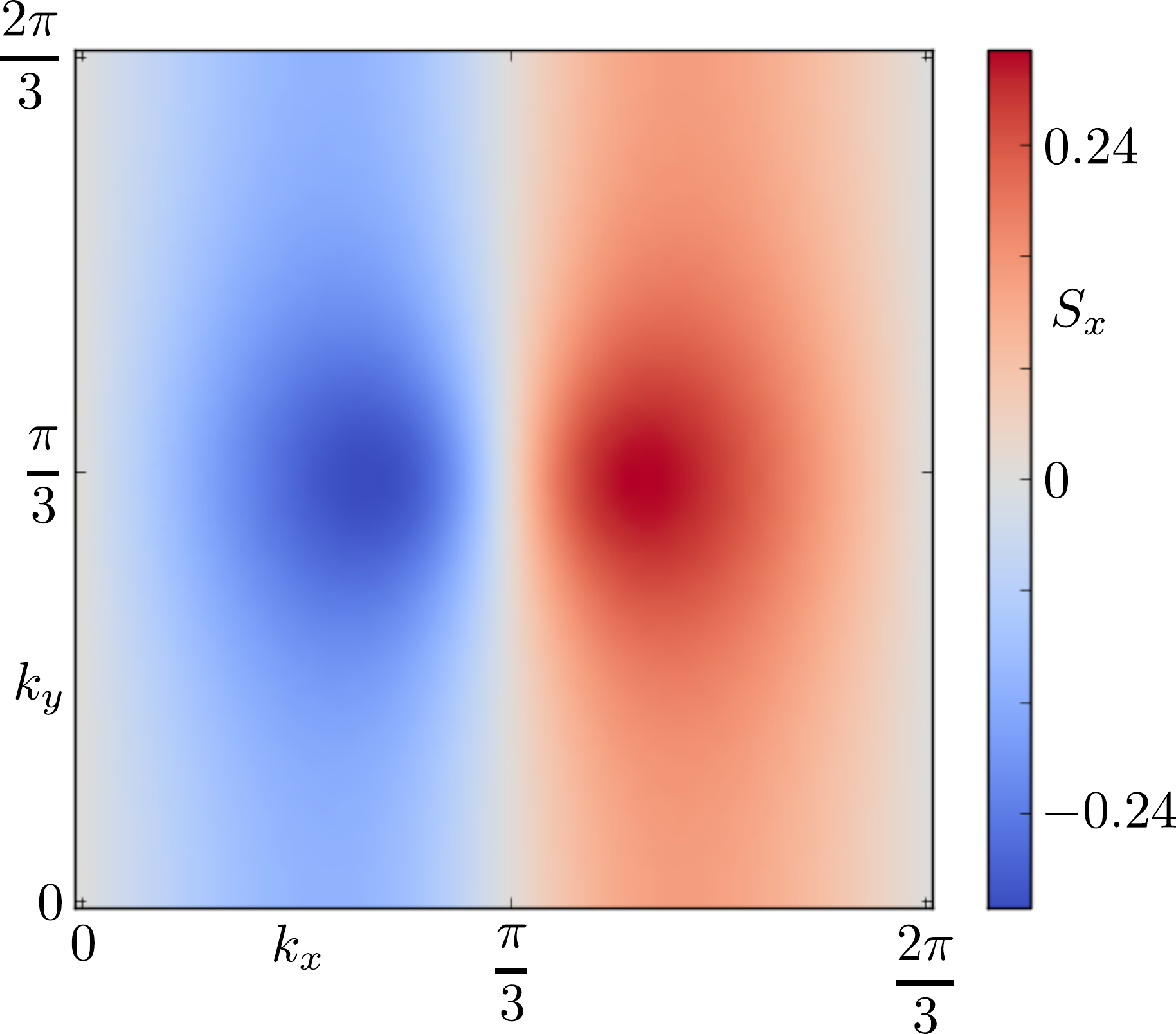}
 \includegraphics[width=0.3\textwidth]{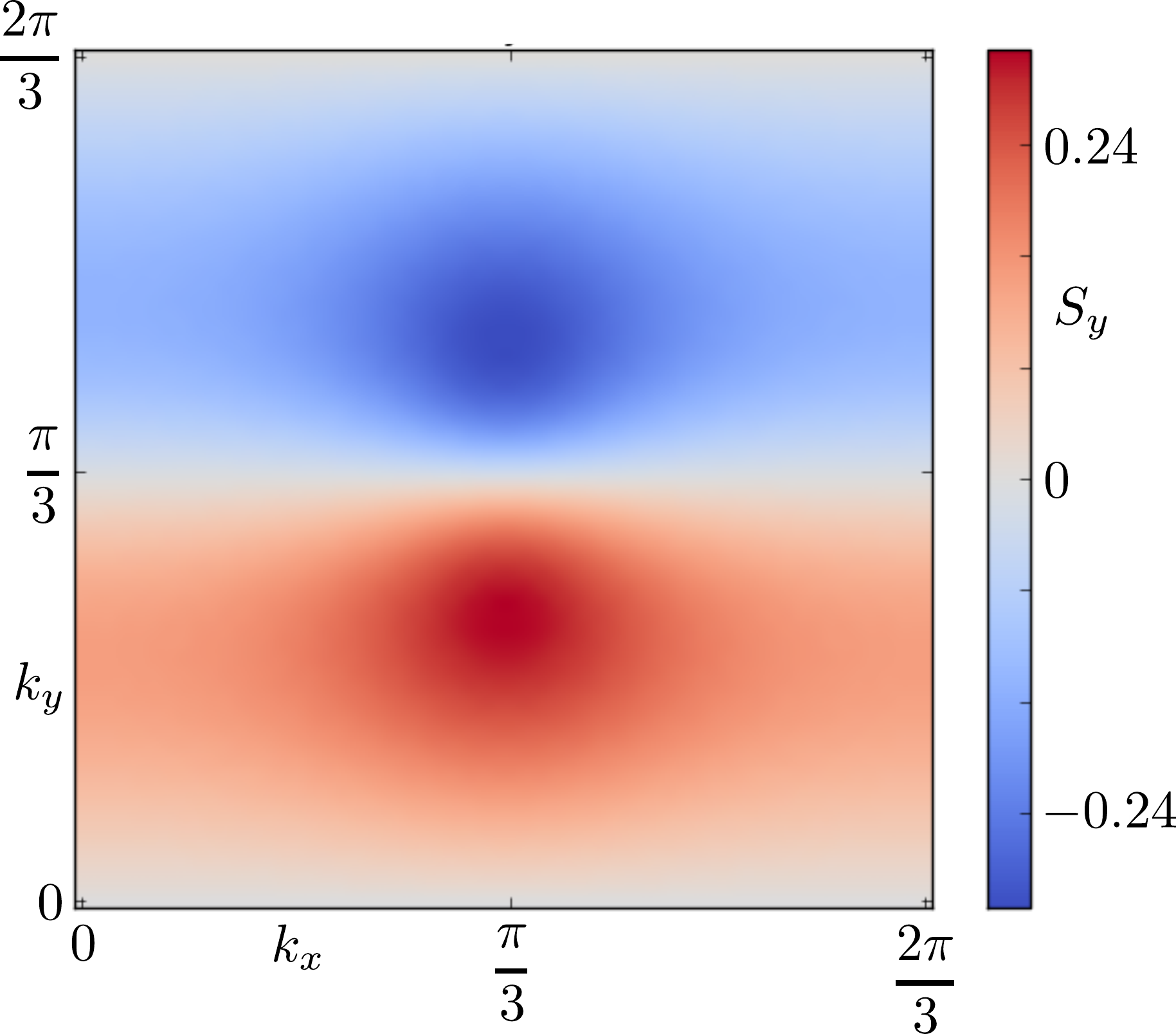}
 \includegraphics[width=0.3\textwidth]{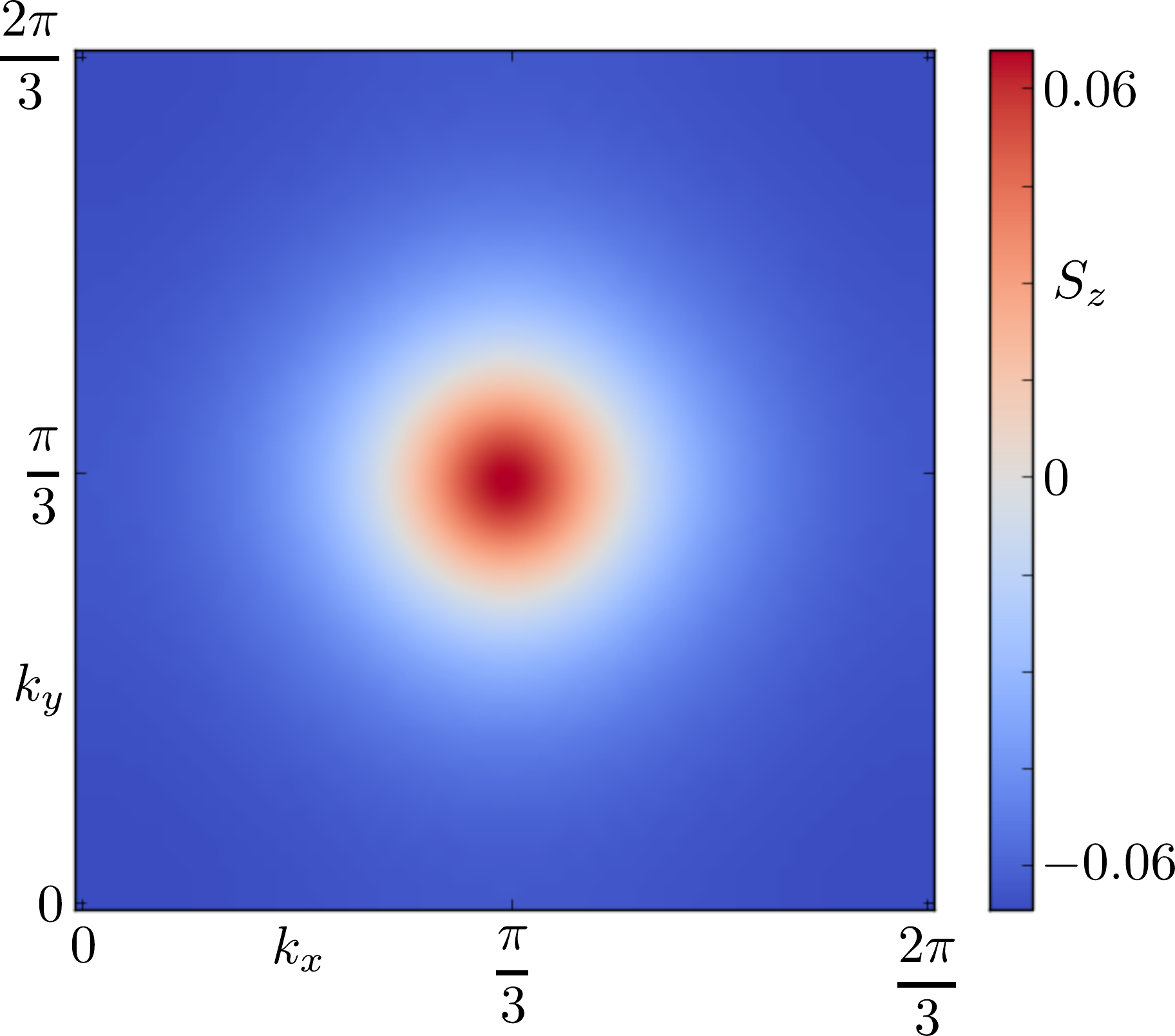}
\caption{(Color online). Comparison between the spin polarization at $q=\pi/5$ (top row), which represents the topological insulating phase with $\mathcal{W}=0$, and $q=2\pi/5$ (bottom row), which describes the second insulating phase with $\mathcal{W}=-1$. The component $S_z$ of the polarization (third column) behaves very differently in the two cases. For $q=\pi/5$ (as generally for $q<\pi/3$), $S_z$ is negative in the whole Brillouin zone (top right panel), whereas for $q=2\pi/5$ (as in general for $\pi/3<q<\pi/2$) the polarization exhibits a skyrmion structure. When $q=2\pi/5$, the skyrmion 
interpolates from $S_z \simeq 0.067$ at $\vec{k}=(0,0)$ to $S_z \simeq -0.067$ at $\vec{k}=(\pi/3,\pi/3)$ (bottom right panel). The $S_x$ and $S_y$ components display opposite signs in the two cases.}
\label{polar}
\end{figure*}

Focusing on the lowest energy band, we define the spin polarization $\vec{S}(\vec{k})$ in the Brillouin zone in terms of a reduced density matrix by 
tracing out the lattice orbital degree of freedom:
\begin{equation}
 \rho(\vec{k})=\sum\limits_{x=0,1,2}  \phantom{\ket{}}_{\rm lat}\bracket{x}{\psi_\lambda(\vec{k})} \bracket{\psi_\lambda(\vec{k})}{x}_{\rm lat},
\end{equation}
where $\rho$ is the $2\times 2$ matrix representing a mixed state for the spin degree of freedom and depending on the quasi-momentum in the Brillouin zone. If only this lowest energy band is filled and the system is not in a semimetal (gapless) phase, the time of flight images, distinguished by their inner spin state and eventually combined with a rotation of the atomic states, allow for the measurement of the observables given by
\begin{equation}
 S_i(\vec{k})=\Tr_{\rm spin}\left[\rho(\vec{k}) \, \sigma_i \right] \, .
\end{equation}
The vector $\vec{S}$ is not normalized due to the fact that $\rho$ describes a mixed state. The behavior of the polarization $\vec{S}$ provides a clear indication of the topological phase transition: for $0<q<\pi/3$ or $2\pi/3<q<\pi$ the component $S_z$ is negative in the whole Brillouin zone, whereas in the intermediate phase at $\pi/3<q<2\pi/3$ (for $q\ne\pi/2$) $S_z$ assumes both negative and positive values (see Fig.~\ref{polar}). At $q=\pi/2$, the polarization always lies in the $\hat{x}-\hat{y}$ plane due to the additional symmetry generated by $\sigma_z$.

The polarization $\vec{S}$ is independent of the gauge choice for the Abelian potential and is periodic with period $\Delta k =2\pi/3$ in both the $\hat{x}$ and $\hat{y}$ directions; thus one has to consider a reduced magnetic Brillouin zone 
with $\vec{k} \in \left[ 0, 2\pi/3 \right) \times \left[ 0, 2\pi/3 \right)$.

Excluding the point at $q=\pi/2$ (where $S_z = 0 $ and our argument fails), the spin polarization along $\hat{z}$ in the phase at $\pi/3<q<2\pi/3$ has a single skyrmion in the reduced Brillouin zone (a configuration in which the spin polarization entirely covers the Bloch sphere once moving in the reduced Brillouin zone). For $\pi/3<q<\pi/2$ this skyrmion is centered at 
$\vec{k}=\left\lbrace \pi/3,\pi/3\right\rbrace$, whereas for 
$\pi/2<q<2\pi/3$ it is centered at $\vec{k}=\left\lbrace 0,0 \right\rbrace$, meaning that the skyrmions appear at the position of the Dirac points at $q=\pi/3$ and $q=2\pi/3$, respectively.

The winding number of the spin polarization in the reduced Brillouin zone (rBZ) provides a good topological invariant that plays the role of an order parameter for these transitions and reflects the presence of a skyrmion in momentum space. In particular this winding number $\mathcal{W}$ is defined as the integral of the spin curvature 
\begin{equation} \label{curv}
c(\vec{k}) = \frac{1}{4\pi} \frac{\vec{S}}{|\vec{S}|^3 }\cdot \left(\partial_{k_x} \vec{S} \times \partial_{k_y} \vec{S} \right) 
\end{equation}
and it reads:
\begin{equation} \label{wind}
 \mathcal{W} = \frac{1}{4\pi} \int_{\rm rBZ} d^2k \, \frac{\vec{S}}{|\vec{S}|^3 }\cdot \left(\partial_{k_x} \vec{S} \times \partial_{k_y} \vec{S} \right) \, ,   
\end{equation}
where $|\vec{S}|$ normalizes the polarization.

The winding number $\mathcal{W}$ is $0$ for $0<q<\pi/3$ and for $2\pi/3<q<\pi$, whereas $\mathcal{W}=-1$ in the other phase (with the exception of the point at $q=\pi/2$). If we want to consider the full Brillouin zone we must multiply this winding number by $3$, and, at the phase transitions, we recover the same discontinuity observed for the Chern number $\mathcal{C}_{-2}$. We find that $\mathcal{C}_{-2}=3\mathcal{W}+1$.

The discrepancy between the two indices cannot be directly related to a contribution to the winding number from the lattice degree of freedom: if we trace out the spin degree of freedom instead of the lattice one and try to calculate, with an analogous procedure, the polarization obtained by the new $3\times 3$ reduced density matrix, its winding number is constantly zero and does not affect the difference $\mathcal{C}_{-2}-3\mathcal{W}$. This unequivalence can be understood even in the absence of non-Abelian flux: a net magnetic field implies that relevant correlators are non-local in momentum space \cite{yoshioka}.

At the experimental level one may be able to estimate $\mathcal{W}$ through 
a discretization of formula \eqref{wind}, based on the division of the time of flight images in small domains, corresponding to a discretization of the Brillouin zone \cite{alba11,goldman13}. It is important to realize that, while experimental time of flight imaging can be well approximated by a continuous set of data, a discretized approach is relevant for analyzing the signal which can be obtained from finite size lattices. The following discussion can therefore be understood as modeling either limitations of the measurement procedure or finite size effects in realistic setups.

In order to illustrate the protocol to compute $\mathcal{W}$, let us divide the reduced Brillouin zone into $L \times L$ plaquettes that we label by discrete  quasimomenta $(p_x,p_y)$, in units of $\delta=\frac{2\pi}{3L}$. For each plaquette it is possible to estimate the vector $\vec{S}(p_x,p_y)$ from the experimental data (eventually repeating the experiments many times to acquire a sufficient statistics). The spin curvature $c(\vec{k})$ in \eqref{curv} 
can be approximated by:
\begin{widetext}
\begin{multline} \label{approx}
 2\pi \, c(p_x,p_y) = \arctan\left[ \frac{  \vec{N}(p_x,p_y)\cdot\left(\vec{N}(p_x+\delta,p_y)\times\vec{N}(p_x,p_y+\delta) \right) }{1+\vec{N}(p_x,p_y)\cdot\vec{N}(p_x+\delta,p_y)+\vec{N}(p_x,p_y+\delta)\cdot\vec{N}(p_x+\delta,p_y)+\vec{N}(p_x,p_y)\cdot\vec{N}(p_x,p_y+\delta)}\right] +\\ 
+  \arctan\left[ \frac{\vec{N}(p_x+\delta,p_y+\delta)\cdot\left(\vec{N}(p_x,p_y+\delta)\times\vec{N}(p_x+\delta,p_y) \right) }{1+\vec{N}(p_x+\delta,p_y+\delta)\cdot\vec{N}(p_x+\delta,p_y)+\vec{N}(p_x,p_y+\delta)\cdot\vec{N}(p_x+\delta,p_y)+\vec{N}(p_x+\delta,p_y+\delta)\cdot\vec{N}(p_x,p_y+\delta)}\right] \, ,
\end{multline}
\end{widetext}
where $\vec{N}$ is the normalized polarization vector $\vec{N}=\frac{\vec{S}}{|\vec{S}|}$.

The two terms in \eqref{approx} correspond to (half) the solid angles \cite{angle} defined respectively by the (ordered) triplets of vectors 
$$\left\lbrace \vec{N}(p_x,p_y),\vec{N}(p_x+\delta,p_y), \vec{N}(p_x,p_y+\delta) 
\right\rbrace$$ 
and 
$$\left\lbrace \vec{N}(p_x+\delta,p_y+\delta),\vec{N}(p_x,p_y+\delta), 
\vec{N}(p_x+\delta,p_y) \right\rbrace.$$ 
By adopting this approximation, $\mathcal{W}$ results from the sum of all 
the $c(p_x,p_y)$'s in the rBZ:
\begin{equation} \label{appr}
 \mathcal{W}\sim \sum\limits_{(p_x , p_y)\in \,\rm{rBZ}} c(p_x , p_y) \, .
\end{equation}

\begin{figure*}[htb]
 \includegraphics[width=\linewidth]{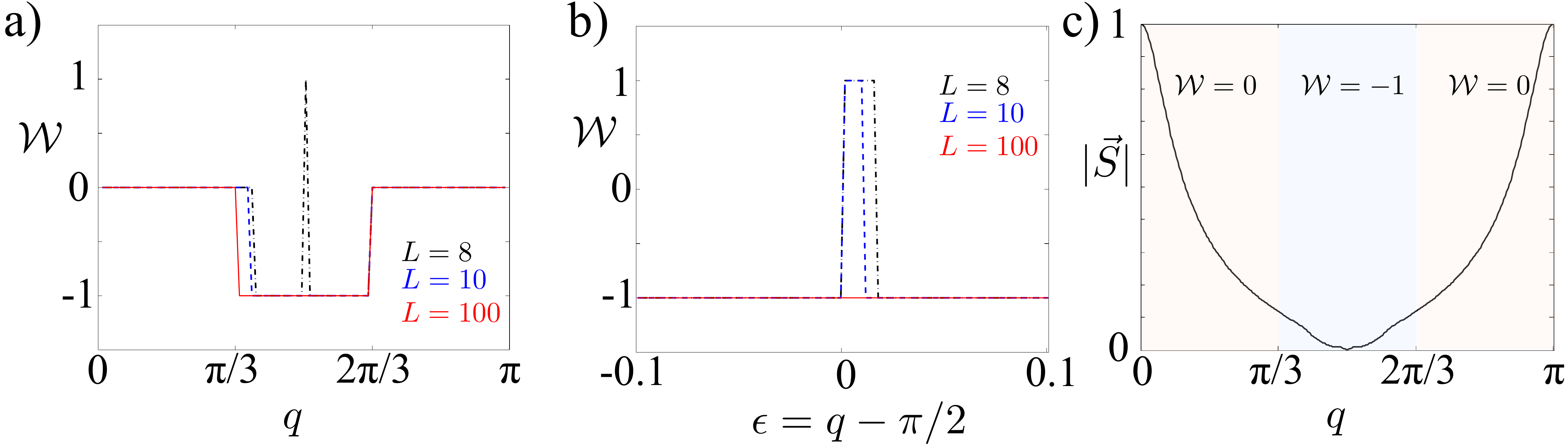} 
\caption{(Color online) Effects of discretization on the relative error in the computation of the winding number. \textit{a):} Estimated spin winding number $\mathcal{W}$ for various values of the discretization $L$: $L=100$ (solid red line), $L=10$ (dashed blue line) and $L=8$ (dashed-dotted black line). The finer grid ($L=100$) reproduces almost exactly the theoretical value, whereas coarser grids feature deviations at the transition points (mainly at $q=\pi/3$ because of the relative positions of the grid and the skyrmion) plus a significative error at the maximally entangled $q=\pi/2$ point. \textit{b):} Winding number computed at $q=\pi/2 + \epsilon$ for different discretizations. \textit{c):} The minimum of the norm of the polarization vector $|\vec{S}|$ as a function of $q$.  This minimum is reached in the skyrmionic region, showing that the detection of the non-trivial winding number requires more statistical power. In particular, at $q=\pi/2$ the norm vanishes at the skyrmion core, rendering the measurement protocol useless at that particular point.}
\label{Errors}
\end{figure*}

Since the obtained value for $\mathcal{W}$ is generally not an integer, 
as it must be, the relative error is easily deducible and it 
is smaller than $1\cdot10^{-3}$ for a $5 \times 5$ 
discretization at $q=\pi/5$ and $q=\pi/4$, corresponding to the 
trivial phase where $\mathcal{W}=0$; the error is also small 
($<1\cdot10^{-3}$) at points in the non-trivial phase away 
from the transition, such as $q=2\pi/5$. However, 
there are a couple of error-prone regions where a finer 
discretization of the reduced Brillouin zone helps to correctly 
determine the topological nature of the state. In particular 
the accuracy decreases both around the phase transitions, 
where the skyrmion might be overlooked by a coarse discretization 
(see Fig.~\ref{Errors}a), and close to $q=\pi/2$ (see Fig.~\ref{Errors}b), 
where the partial density matrix approach fails because 
the orbital and pseudospin degrees of freedom become maximally 
entangled and the $S_z$ component of the spin vanishes. 
The main contribution to the total winding number is 
indeed obtained from the central region of each skyrmion, 
where the gradient of $\vec{N}$ is larger. 
Therefore one has to ensure sufficient sampling close to the 
skyrmion core, in order to avoid unwanted large discretization errors.

We also note that, in the presence of a skyrmion, the norm of 
the polarization is much smaller than unity (see Fig.~\ref{Errors}c) 
due to a strong mixing 
with the lattice degree of freedom: 
for example at $q=2\pi/5$ the component $S_z$ in the north and south poles 
appearing at $\vec{k}=(0,0)$ and $\vec{k}=(\pi/3,\pi/3)$ 
is approximately $\pm 0.067$ (see Fig.~\ref{polar}). 
Thus the difference in the number of atoms between the two species at the center of the skyrmion is of the order of 10\%. This example is relevant since $q=2\pi/5$ is close to the optimal point in which $S_z$ is maximized in the center and there we find the maximal extension of the core of the skyrmion 
(defined as the neighborhood of the north pole where $S_z>0$). In general the dimension of the core is quite reduced and, in order to avoid excessive discretization errors, we adopted the approximation in Eq. (\ref{approx}), which is finer than the one used in previous works \cite{alba11,goldman13} 
and better fulfills the accuracy requirements of our model.

As a final remark we point out that the detection of the phase transition 
by the spin winding number works if only the lowest band is occupied. In the semimetal phase also the second band contributes to the measured spin winding number, thus spoiling the final result (in \cite{goldman13} the behavior of the spin winding number is investigated also in the semimetal regime where two different bands are involved).

\section{Edge states in the topological phases}
The difference between the two topological insulating phases manifests itself also in the edge state structure. We already mentioned that the edge mode propagation direction of the lowest energy band changes across the topological phase transition. Therefore the techniques presented in Refs.~\cite{goldman13b,goldman13d} can be applied to distinguish the two topological phases by measuring their respective edge chiralities. 

\begin{figure}[htb]
 \includegraphics[width=0.45\textwidth]{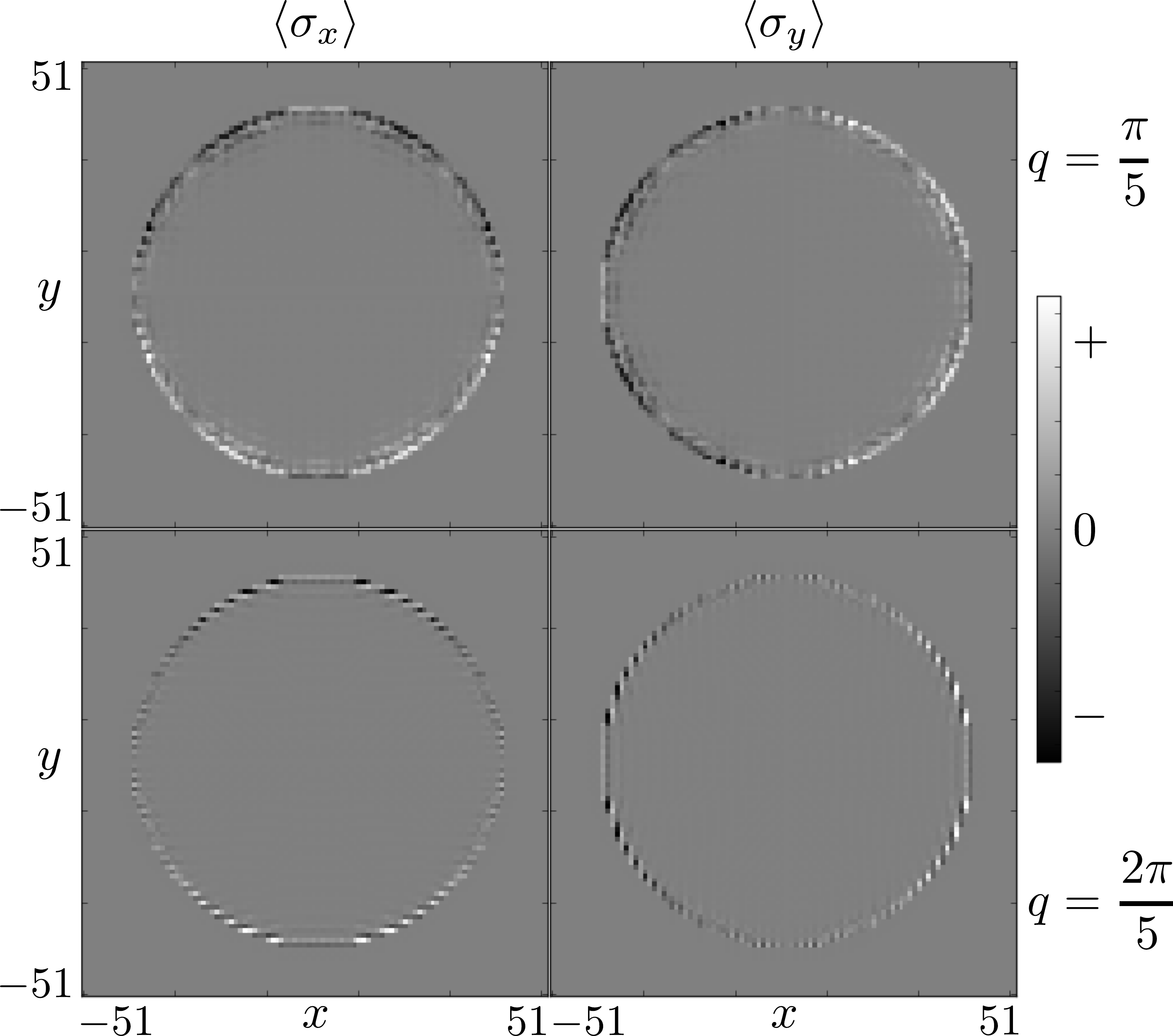} 
\caption{In-plane spin polarization of the edge modes of the lowest energy band in the two different topological phases at $q=\pi/5$ and $q=2\pi/5$. 
The average spin densities along $x$ and $y$ are shown for a single edge state in the first energy gap, for a hard-wall potential with a $40$ site radius. The top and bottom rows show the results for $q=\pi/5$ and $q=2\pi/5$. The spin distribution is qualitatively the same in the two cases.}
\label{polar_edge}
\end{figure}

The real space, edge mode spin polarization instead does not allow us to detect the phase transition, since it presents qualitatively the same behavior in the two phases. In Fig.~\ref{polar_edge} we plot the in-plane spin polarization of one of the states in the first band gap at $q=\pi/5$ and $q=2\pi/5$, corresponding to the two different insulating topological phases. 
Adopting a hard-wall potential to define the edges of the system, we calculated the densities $\langle \sigma_x \rangle = 2\Re[\Psi_{\uparrow}^*\Psi_{\downarrow}]$ and $\langle \sigma_y \rangle = 2 \Im[\Psi_{\uparrow}^*\Psi_{\downarrow}]$ as functions of the position 
($\Psi_\uparrow$ and $\Psi_\downarrow$ are the amplitudes of the edge mode 
wave function with a defined $\hat{z}$ component of the spin). In this way we can analyze the $x$ and $y$ components of the spin along the edge. The behavior of the spin polarization along the edge is similar in both phases. Since the chirality in the two cases is inverted, this implies that also the helicity $\vec{k}\cdot \vec{\sigma}$ changes sign. This behavior is consistent with the one in the bulk (see the first two columns of Fig.~\ref{polar}), where the in-plane spin polarization as a function of the quasi-momentum is inverted across the phase transition.

The different topological properties of the two phases lead to a doubling of the edge states across the transition, so the analysis of the mass density $|\Psi_\uparrow|^2+|\Psi_\downarrow|^2$ of the states appearing in the first gap may help to distinguish the two topological phases.

\begin{figure}[ht]
 \includegraphics[width=0.475\textwidth]{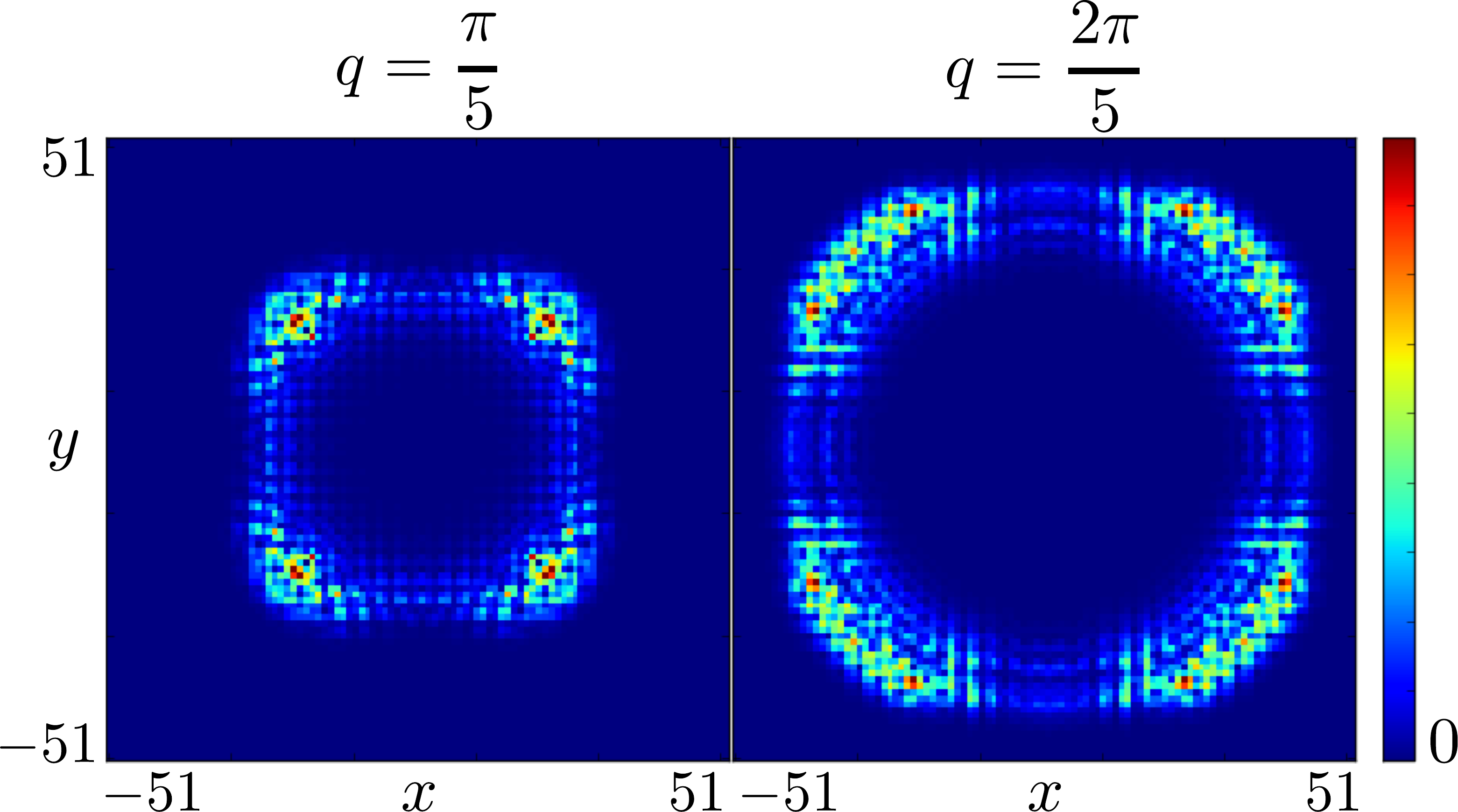} 
 \caption{(Color online). Total mass density $|\Psi_\uparrow|^2+|\Psi_\downarrow|^2$ for two edge states of the lowest energy band at $q=\pi/5$ and $q=2\pi/5$ and under a harmonic trapping potential. A typical interference pattern breaking the axial symmetry appears. The system size is $L=102$ and the trapping radius is $R=2L/5$.}
\label{harm}
\end{figure}

As a realistic setup we consider first a harmonic trapping potential of the form  $V_0 (r/R)^2$, where $V_0$ is proportional to the trapping frequency (set to the value $V_0=t$), with $r$ being the distance from the center of the lattice and $R$ being a characteristic radius for the trapping. Typical profile densities for edge states at $q=\pi/5$ and $q=2\pi/5$ are shown in Fig.~\ref{harm}. Also for $q=2\pi/5$, where the edge distribution is broader, it is difficult to distinguish the presence of two edge states, mainly due to their interference pattern.

To better evaluate the edge structure we considered a different potential geometry with a trapping which acts only along the $y$ direction, by adding an on-site energy $V(y) = V_{0;y} (y/R)^\gamma $ ($V_{0;y}=t$). The extension and position of the horizontal edge modes depend on the choice of $\gamma$ (the hard-wall boundary is recovered for $\gamma \to \infty$), whereas the vertical edge modes appear on the boundaries of the system at $x=\pm L/2$ and do not depend on the chosen trapping potential since they are always defined by open boundary conditions.

\begin{figure}[ht]
 \includegraphics[width=0.475\textwidth]{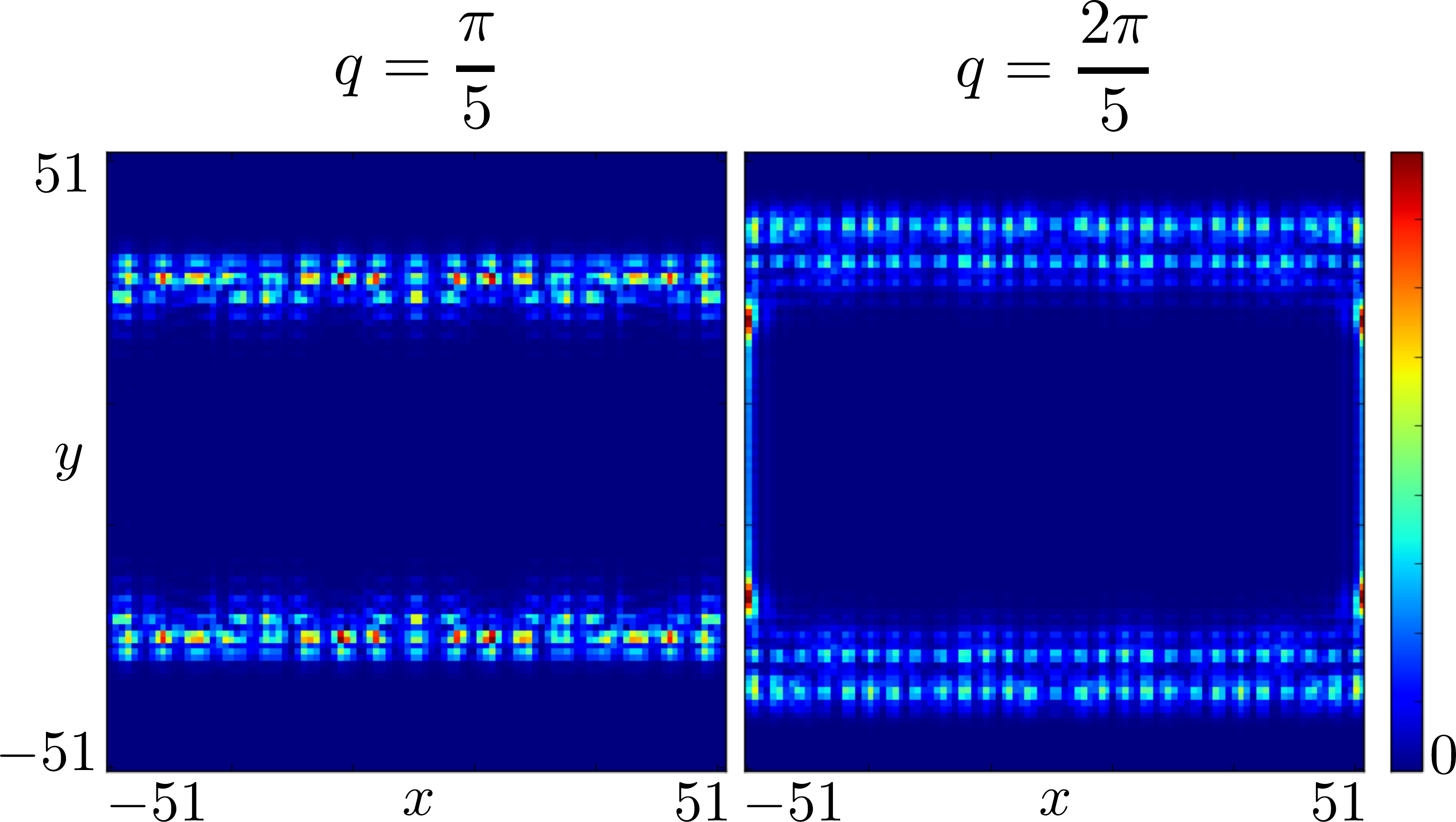} 
 \caption{(Color online). Total density $|\Psi_\uparrow|^2+|\Psi_\downarrow|^2$ of an eigenstate with energy lying in the first band gap in a lattice with harmonic potential along the $y$ direction for $q=\pi/5$ (left) and $q=2\pi/5$ (right). In the upper and lower edges of the right panel, two edge modes are clearly distinguishable. On the vertical edges the two modes can no longer be distinguished due to their stronger localization dictated by the hard-wall condition on the boundary.}
\label{harm2pi5}
\end{figure}

Adopting the harmonic potential the difference between a single edge mode 
in the $q<\pi/3$ regime and two edge modes for $ \pi/3 < q< 2 \pi/3$ 
becomes more evident in this geometry. In Fig.~\ref{harm2pi5} we plot 
the total wave function density of one eigenstate with energy lying in the first band gap for both $q=\pi/5$ and $q=2\pi/5$. The structure of the corresponding edge modes is clearly visible: on the horizontal edges, determined by the harmonic potentials, a single edge mode appears for $q=\pi/5$, whereas two separated modes can be distinguished for $q=2\pi/5$. All these horizontal modes present a typical oscillating density. Along the vertical boundaries, instead, the open boundary condition, corresponding to a hard wall potential, implies a stronger localization, which does not allow us to distinguish the two phases since the two modes at $q=2\pi/5$ overlap.

\begin{figure*}[ht]
 \includegraphics[width=0.95\textwidth]{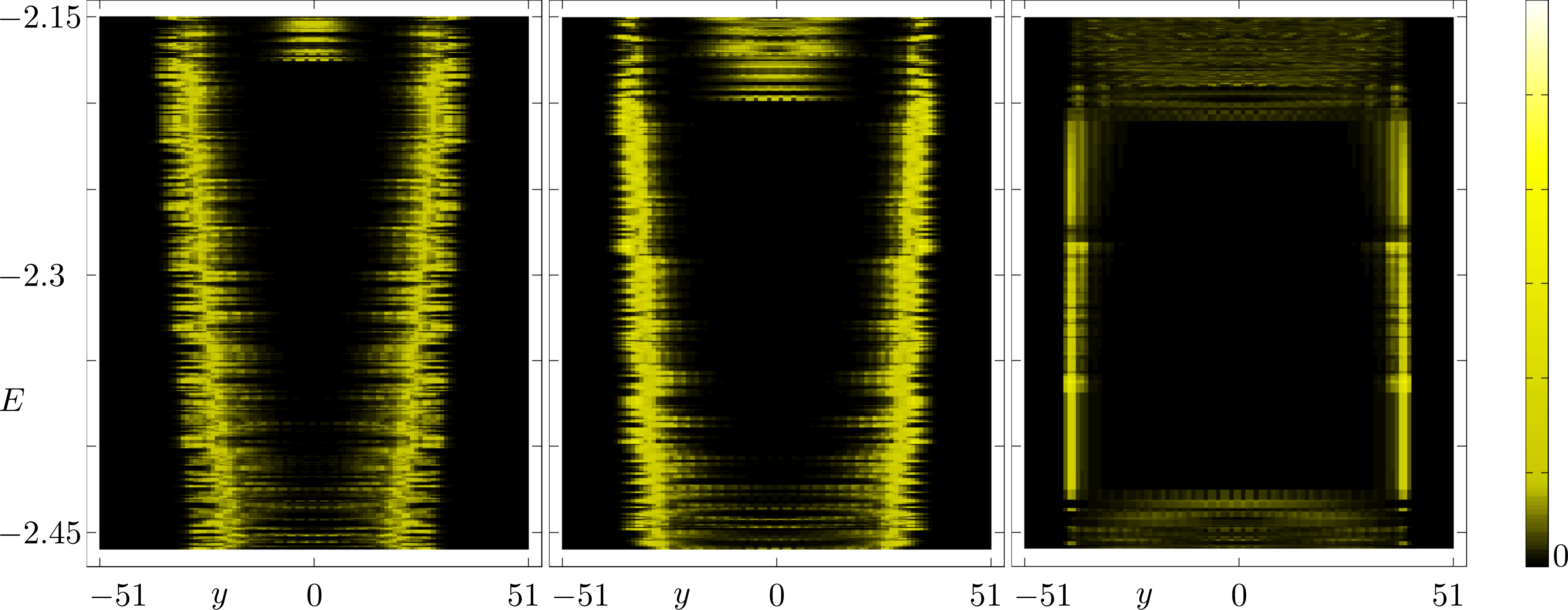}
\caption{(Color online). Density distribution of the edge states in the first gap, shown for $q=\pi/5$ in a square system of size $102 \times 102$, as a function of energy and position. The dark central zone in each panel is the band gap separating the two lower bands (central yellow regions at the top and the bottom). The edge states are the yellow lateral lines.  The left panel represents a system with a harmonic trapping potential, the middle one refers to a quartic potential, and the right one refers to a hard-wall potential. The trappings are only along the $y$ direction. The densities are calculated by averaging the total wave function density along the $x$ direction in the central region of the system (for $-30\le x \le 30$), far from the vertical edges.}
\label{trapdensity}
\end{figure*}

To evaluate the effect of the trapping potential on the edge modes in Fig.~\ref{trapdensity} we plot the wave function density $\left| \Psi (y) \right|^2$ of all states appearing between the two lowest bands at $q=\pi/5$ as a function of their energy. All the numerical results in this section were obtained using the Kwant software package \cite{kwant,kwant2}. 

We considered harmonic, quartic and hard-wall trapping potentials. In order to avoid the influence of both the interference pattern, strongly dependent on the energy of the eigenstates, and the vertical edge modes, we averaged the density of the wave functions along the $x$ direction in the central region of the lattice, far from the vertical edges. Analogously to the case of topological insulators of spinless fermions \cite{goldman13b,goldman13d}, by increasing $\gamma$ the edge states become more and more localized. A similar behavior characterizes also the other phase where the presence of two different edge modes makes the edge density broader.

\section{Conclusions}
We analyzed a model of two-component fermions on a square optical lattice with tunneling amplitudes determined by a non-Abelian gauge potential with 
both a magnetic field and a translationally invariant SU(2) non-Abelian term. 
By choosing a magnetic flux different from $\Phi_0/2$ we considered a setup with broken time-reversal symmetry and showed that the non-Abelian term drives the system across topological phase transitions. In particular we investigated the case of $\Phi=\Phi_0/3$, where the system presents six energy bands, and discussed its phase diagram at filling $1/3$, which is characterized by a topological semimetal and by two different topological insulating phases, with edge states differing in number and in propagation direction across the topological phase transition.

The topological insulating phases can be distinguished both by looking directly at the edge states of the system (with accuracy depending on the trapping potential) and by the spin polarization in momentum space, estimating the spin winding number from time of flight absorption images. The latter method provides a clear diagnostics for the characterization of the topological phase transition even though the spin is not a conserved quantity of the system.

The single-particle properties of this system may constitute the basis for 
the study of topological phase transitions also in the presence of interactions. In particular, in the limit $q\to \pi/2$ the lowest band in our model has an extremely flat profile which constitutes a fundamental requirement to engineer Chern insulators, able to mimic fractional quantum Hall physics \cite{sondhi13}, even though in this limit measuring the spin winding number would become more difficult. 

Furthermore, introducing a repulsive interaction may lead to the regime of topological Mott insulators \cite{dauphin12} where the presence of the spin degree of freedom may enrich the phase diagram of such interacting models, also due to the possibility of introducing spin-dependent interactions. Finally, the interactions may also modify and merge the Dirac points in the system \cite{fu13}.

\vspace{0.5cm}

\acknowledgements
We warmly thank Leonardo Mazza and Fabio Mezzacapo for useful and stimulating discussions. 
This work was supported by the Dutch Science Foundation NWO/FOM and by an 
ERC Advanced Investigator Grant. A.T. acknowledges support from the STREP MatterWave. E.A. acknowledges funding from FPU 2009-1761.
L.L. also acknowledges a grant from Banco de Santander and 
financial support from ERDF.


\begin{thebibliography}{XX}

\bibitem{hasankane} M. Z. Hasan and C. L. Kane, Rev. Mod. Phys. \textbf{82}, 3045 (2010).

\bibitem{zhang11} X.-L. Qi and S.-C. Zhang, Rev. Mod. Phys. \textbf{83}, 1057 (2011).

\bibitem{dalibard11} J. Dalibard, F. Gerbier, G. Juzeliunas, and P. \"Ohberg
Rev. Mod. Phys. \textbf{83}, 1523 (2011). 

\bibitem{Lewensteinbook}
M. Lewenstein, A. Sanpera, and V. Ahufinger, 
{\em Ultracold atoms in optical lattices: simulating quantum many-body systems} 
(Oxford, Oxford University Press, 2012).

\bibitem{goldman10} N. Goldman, I. Satija, P. Nikolic, A. Bermudez, M. A. Martin-Delgado, 
M. Lewenstein, and I. B. Spielman, Phys. Rev. Lett. \textbf{105}, 255302 (2010).

\bibitem{Tarr} L. Tarruell, D. Greif, T. Uehlinger, G. Jotzu, and T. Esslinger, Nature \textbf{483}, 302 (2012).

\bibitem{alba11} E. Alba, X. Fernandez-Gonzalvo, J. Mur-Petit, J. K. Pachos, and J. J. Garcia-Ripoll, 
Phys. Rev. Lett. \textbf{107}, 235301 (2011).

\bibitem{goldman13} N. Goldman, E. Anisimovas, F. Gerbier, P. \"Ohberg, I. B. Spielman, 
and G. Juzeliunas, New J. Phys. \textbf{15}, 013025 (2013).

\bibitem{haldane} F. D. M. Haldane, Phys. Rev. Lett. \textbf{61}, 2015 (1988).

\bibitem{affleck} I. Affleck and J. B. Marston, Phys. Rev.  B \textbf{37}, 3774 (1988).

\bibitem{bernevig} B. A. Bernevig, D. Giuliano, and R. B. Laughlin, Ann. Phys. \textbf{311}, 
182 (2004).

\bibitem{manes} J. L. Manes, Phys. Rev. B \textbf{85}, 155118 (2012). 

\bibitem{hasegawa} Y. Hasegawa, J. Phys. Soc. Jpn. \textbf{59}, 4384 (1990).

\bibitem{laughlin} R. B. Laughlin and Z. Zou, Phys. Rev. B \textbf{41}, 664 (1990).

\bibitem{LMT} L. Lepori, G. Mussardo, and A. Trombettoni, Europhys.
Lett. \textbf{92}, 50003 (2010).

\bibitem{MLT} G. Mazzucchi, L. Lepori, and A. Trombettoni, J. Phys. B: At. Mol. Opt. Phys. {\bf46} 134014 (2013).

\bibitem{lewenstein13} N. Goldman, F. Gerbier, and M. Lewenstein,  J. Phys. B \textbf{46}, 134010 (2013).

\bibitem{osterloh05} K. Osterloh, M. Baig, L. Santos, P. Zoller, 
and M. Lewenstein, Phys. Rev. Lett. \textbf{95}, 010403 (2005).

\bibitem{goldman09a} N. Goldman, A. Kubasiak, A. Bermudez, P. Gaspard, M. Lewenstein, and M. A. Martin-Delgado, Phys. Rev. Lett. {\bf 103}, 035301 (2009).

\bibitem{goldman09b} N. Goldman, A. Kubasiak, P. Gaspard, and M. Lewenstein, Phys. Rev. A {\bf{79}}, 023624 (2009).

\bibitem{goldman12} N. Goldman, W. Beugeling, and C. Morais Smith,  Europhys. Lett. \textbf{97}, 23003 (2012).

\bibitem{morais} W. Beugeling, N. Goldman, and C. Morais Smith, Phys. Rev. B \textbf{86}, 075118 (2012).

\bibitem{morais2} W. Beugeling, J. C. Everts, and C. Morais Smith, Phys. Rev. B \textbf{86}, 195129 (2012).

\bibitem{lin09} Y. J. Lin, R. L. Compton, K. Jimenez-Garcia, J. V. Porto, and
I. B. Spielman, Nature \textbf{462}, 628 (2009).

\bibitem{jaksch03} D. Jaksch and P. Zoller,  New J. Phys. \textbf{5}, 56 (2003).

\bibitem{mazza12} L. Mazza, A. Bermudez, N. Goldman, M. Rizzi, M. A. Martin-Delgado, 
and M. Lewenstein, New J. Phys. \textbf{14}, 015007 (2012).

\bibitem{aidelsburger11} M. Aidelsburger, M. Atala, S. Nascimbene, S. Trotzky, Y.-A. Chen, and 
I. Bloch, 2012 Phys. Rev. Lett. \textbf{107} 255301 (2012).

\bibitem{jimenez12} K. Jimenez-Garcia, L. J. LeBlanc, R. A. Williams, M. C. Beeler, A. R. 
Perry, and I. B. Spielman, Phys. Rev. Lett. \textbf{108}, 225303 (2012).

\bibitem{hauke12} P. Hauke, O. Tieleman, A. Celi, C. \"Olschl\"ager, J. Simonet, J. Struck,
M. Weinberg, P. Windpassinger, K. Sengstock, M. Lewenstein, and A. Eckardt,  
Phys. Rev. Lett. \textbf{109} 145301 (2012).

\bibitem{brantut12} J.-P. Brantut, J. Meineke, D. Stadler, S. Krinner, and T. Esslinger, 
Science \textbf{337}, 1069 (2012).

\bibitem{goldman13b} N. Goldman, J. Beugnon, and F. Gerbier, Phys. Rev. Lett. 
\textbf{108}, 255303 (2012); Eur. Phys. J. Special Topics \textbf{217}, 135 (2013).

\bibitem{goldman13d} N. Goldman, J. Dalibard, A. Dauphin, F. Gerbier, M. Lewenstein, P. Zoller, 
and I. B. Spielman, PNAS {\bf 110}, 6736 (2013).

\bibitem{leblanc12} L. J. LeBlanc, K. Jim\'enez-Garc\'ia, R. A. Williams, M. C. Beeler, A. R. Perry, W. D. Phillips, and I. B. Spielman, PNAS \textbf{109}, 10811 (2012).

\bibitem{barberan13} H. Pino, E. Alba, J. Taron, J. J. Garcia-Ripoll and N. Barber\'an, 
Phys. Rev. A  \textbf{87}, 053611 (2013).

\bibitem{dauphin13} A. Dauphin and N. Goldman, arXiv:1305.3872                                    

\bibitem{cooper12} H. M. Price and N. R. Cooper, Phys. Rev. A \textbf{85}, 033620 (2012); 
arXiv:1306.4796 

\bibitem{lee13} X.-J. Liu, K. T. Law, T. K. Ng, and P. A. Lee, arXiv:1306.5223

\bibitem{satija11} E. Zhao, N. Bray-Ali, C. J. Williams, I. B. Spielman, and I. I. Satija, 
Phys. Rev. A \text{bf 84}, 063629 (2011).

\bibitem{troyer13} L. Wang, A. A. Soluyanov, and M. Troyer, Phys. Rev. Lett. 
\textbf{110}, 166802 (2013).

\bibitem{burrello10} M. Burrello and A. Trombettoni, Phys. Rev. Lett. \textbf{105}, 125304 (2010).

\bibitem{burrello11} M. Burrello and A. Trombettoni, Phys. Rev. A \textbf{84}, 043625 (2011).

\bibitem{palmer11} R. N. Palmer and J. K. Pachos, New J. Phys. \textbf{13}, 065002 (2011).

\bibitem{grass12} T. Grass, B. Juli\'a-D\'iaz, N. Barber\'an and M. Lewenstein, Phys. Rev. A \textbf{86}, 021603(R) (2012).

\bibitem{komineas12} S. Komineas and N. R. Cooper, Phys. Rev. A \textbf{85}, 053623 (2012).

\bibitem{grass13} T. Grass, B. Juli\'a-D\'iaz, M. Burrello and M. Lewenstein, J. Phys. B: At. Mol. Opt. Phys. \textbf{46}, 134006 (2013).

\bibitem{ludwig08} A. P. Schnyder, S. Ryu, A. Furusaki, and A. W. W. Ludwig, 
Phys. Rev. B \textbf{78}, 195125 (2008). 

\bibitem{radic12} J. Radic, A. Di Ciolo, K. Sun, and V. Galitski, 
Phys. Rev. Lett. \textbf{109}, 085303 (2012). 

\bibitem{LandauStat2} E. M. Lifschitz and L. P. Pitaevskii, 
{\em Statistical Physics, Part 2} (Pergamon Press, 1980).

\bibitem{gerbier10} F. Gerbier and J. Dalibard, New J. Phys. \textbf{12}, 033007 (2010).

\bibitem{altlandzirnbauer} A. Altland and M. R. Zirnbauer, Phys. Rev. B \textbf{55}, 1142 (1997).

\bibitem{kitaev09} A. Kitaev, AIP Conf. Proc. {\bf 1134}, 22 (2009).

\bibitem{ludwig10} S. Ryu, A. P. Schnyder, A. Furusaki, and A. W. W. Ludwig, New J. Phys. {\bf 12}, 065010 (2010).

\bibitem{helical} X.-L. Qi, T. L. Hughes, S. Raghu, and S.-C. Zhang,
Phys. Rev. Lett. \textbf{102}, 187001 (2009). 

\bibitem{goldman07} N. Goldman and P. Gaspard, Europhys. Lett. \textbf{78}, 60001 (2007). 

\bibitem{note1}
The symmetry class is A with the exception of the particular case with 
$q=\pi/2$ and half-filling 
($\mu=0$) when $\sigma_z$ generates a further discrete unitary symmetry, 
$\sigma_z H(\vec{k}) \, \sigma_z=-H(\vec{k})$, and the symmetry class 
becomes the topologically trivial chiral unitary class (AIII).

\bibitem{tknn} D. J. Thouless, M. Kohmoto, M. P. Nightingale, and M. den Nijs, 
Phys. Rev. Lett. \textbf{49}, 405 (1982).

\bibitem{kwant} The {\sc kwant} 
package, developed by A. R. Akhmerov, C. W. Groth, X. Waintal, and M. Wimmer, is available at
{\tt www.kwant-project.org}.

\bibitem{kwant2} C. W. Groth, M. Wimmer, A. R. Akhmerov and X. Waintal, arXiv:1309.2926 (2013).

\bibitem{pachos13} J. K. Pachos, E. Alba, V. Lahtinen, and 
J. J. Garcia-Ripoll, Phys. Rev. A \textbf{88}, 013622 (2013).

\bibitem{yoshioka} D. Yoshioka, {\em The quantum Hall effect} (Berlin, Springer-Verlag, 2002). 

\bibitem{angle} A. van Oosterom and J. Strackee, IEEE Trans. Biom. Eng. BME \textbf{30}, 125 (1983).

\bibitem{sondhi13} S. A. Parameswaran, R. Roy, and S. L. Sondhi, arXiv:1302.6606 

\bibitem{dauphin12} A. Dauphin, M. M\"uller, and M. A. Martin-Delgado, 
Phys. Rev. A \textbf{86}, 053618 (2012).

\bibitem{fu13} L. Wang and L. Fu, Phys. Rev A \textbf{87}, 053612 (2013).


\end{thebibliography}
\end{document}